\documentclass[a4paper,11pt]{article}

\usepackage{jheppub} 

\usepackage[T1]{fontenc} 

\newcommand\re {\mathrm{Re}}
\newcommand\im {\mathrm{Im}}

\title{\boldmath Entanglement Entropy of a Scalar Field in a Squeezed State}

\author[a,b]{D. Katsinis,}
\author[b,c]{G. Pastras}
\author[b]{and N. Tetradis}

\affiliation[a]{Instituto de F\'isica, Universidade de S\~ao Paulo,\\ Rua do Mat\~ao Travessa 1371, 05508-090 S\~ao Paulo, SP, Brazil}
\affiliation[b]{Department of Physics, University of Athens,\\Zographou 157 84, Greece}
\affiliation[c]{Laboratory for Manufacturing Systems and Automation, Department of Mechanical Engineering and Aeronautics, University of Patras,\\Rio, Patra 26504, Greece}

\emailAdd{dkatsinis@phys.uoa.gr}
\emailAdd{pastras@lms.mech.upatras.gr}
\emailAdd{ntetrad@phys.uoa.gr}

\abstract{We study the entanglement entropy within a spherical region for a free scalar field in a squeezed state in $3+1$ dimensions. We show that, even for small squeezing, a volume term appears, whose coefficient is essentially independent of the field mass. This is in line with Page's argument that the entanglement entropy in an arbitrary quantum state is proportional to the number of degrees of freedom of the smaller subsystem. It follows that squeezed states can be considered as arbitrary quantum states, in contrast to the ground or coherent states that give rise to entanglement entropy dominated by a term proportional to the area of the entangling surface.}

\begin{document} 
\maketitle
\flushbottom

\section{Introduction}
\label{sec:intro}

A long time ago, in the groundbreaking works \cite{Bombelli:1986rw,srednicki,Sorkin:1984kjy}, it was realized that entanglement entropy in the ground state of scalar field theory obeys an area law. This is an impressive similarity with the entropy of black holes. It is natural to wonder whether, and to what extent, entanglement entropy contributes to the entropy of black holes. Moreover, one may pose the more fundamental question whether gravity is an emergent statistical force, originating from entanglement. These ideas have been explored in \cite{Jacobson:1995ab,VanRaamsdonk:2010pw,Jacobson:2015hqa}. In the context of the AdS/CFT correspondence they obtain a concrete form. In particular, the linearized Einstein's equations around the pure AdS geometry in the bulk can be derived as a holographic manifestation of the first law of entanglement thermodynamics on the boundary \cite{Lashkari:2013koa,Faulkner:2013ica}.

The above concern the vacuum state of a quantum system, which is a very special state. There is no reason to expect that the entanglement entropy in the vacuum state is similar or shares the same characteristics as the entanglement entropy in an arbitrary quantum state. Actually, roughly 30 years ago, Page proved that the bipartite entanglement entropy in an arbitrary quantum state is close to maximal \cite{Page:1993df}, with a value proportional to the number of degrees of freedom of the smaller subsystem. One should keep in mind that the dimensionality of the Hilbert space of each local degree of freedom affects the constant of proportionality. In a scalar quantum field theory, each local degree of freedom has an infinite-dimensional Hilbert space, which essentially implies that this proportionality constant is unbounded. Nevertheless, entanglement entropy still has to scale with the number of degrees of freedom of the smaller subsystem, in other words with its volume. In the context of black hole physics, this line of reasoning leads to the famous Page curve for the entropy of Hawking radiation \cite{Page:1993wv}. 

Entanglement in field theory is a vast subject, for which indicative reviews include \cite{Calabrese:2004eu,Casini:2009sr,Calabrese:2009qy}. The present work is closely related to Srednicki's method \cite{srednicki} for computing the entanglement entropy. This method is applicable to the ground state of the overall system, for which the reduced density matrix is Gaussian and its spectrum can be obtained analytically. This is not true in an arbitrary state of the system, even in simple states such as the energy eigenstates. In our previous work we generalized Srendicki's method to other cases in which the reduced density matrix is Gaussian. For example, by adding a mass term one can derive analytical results by making use of the inverse of the mass as a perturbative parameter \cite{Katsinis:2017qzh}. A similar treatment can be applied to thermal states \cite{Katsinis:2019vhk,Katsinis:2019lis}. The main difference to the ground-state case is that the area law is obeyed by the mutual information, and not by the entanglement entropy. The latter scales with the volume because of the classical thermal correlations between the subsystems. One can also study coherent states. It turns out that the spectrum of the reduced density matrix is identical to that of the ground state \cite{Benedict:1995yp}, and the reduced system is described by an effective quadratic, but explicitly time-dependent, Hamiltonian \cite{Katsinis:2022fxu}. 

The most general Gaussian wave functions, namely those of squeezed states, can be used in order to probe less special quantum regimes of the field theory. Entanglement properties of these states have been studied in a number of fields, such as optics \cite{9784123,PhysRevLett.86.4267,Laurat_2005} and quantum information \cite{Braunstein:2005zz}. It is interesting to investigate whether these states can be considered arbitrary enough for Page's arguments to be applicable. In our latest work \cite{Katsinis:2023hqn}, see also \cite{Boutivas:2023ksg}, we applied Srednicki's method to these states for a field theory 
in $1+1$ dimensions. The calculation of the reduced density matrix is straightforward. However, the specification of its spectrum requires some non-trivial modifications, because the matrices appearing in the exponent of the resulting expression are not real. We showed that for very large squeezing the entanglement entropy contains a volume term that is proportional to the squeezing parameter. An alternative treatment of squeezed states is presented in \cite{Bianchi:2015fra,Adesso:2014}. Similar calculations can also be performed for fermionic systems \cite{Bianchi:2021lnp}, and it was shown that the mean entanglement entropy indeed approaches the Page curve.

Srednicki's method is essentially a quantum mechanics calculation and not a quantum field theory one, i.e. it can be applied to systems with countable degrees of freedom. This forces us to work with a discretized version of the scalar field theory and not with its original continuous formulation. For reasons that we explain in section \ref{sec:regularization}, it is preferable to have a smooth entangling surface, with the
 simplest choice being that of a sphere. Since we are interested in tracing out a spherically symmetric region, it is natural to discretize the degrees of freedom on a lattice of spherical shells. The basic drawback of such a discetization is that the degrees of freedom are not distributed uniformly. This arises because we impose an UV cutoff on the wavelength of normal modes in the radial direction, but not in the angular ones. Since Page's argument makes use of the number of degrees of freedom, one should make sure that this number scales appropriately with the radius of the subsystem. In order to avoid this complication, in \cite{Katsinis:2023hqn} we restricted our analysis to $(1+1)$-dimensional field theory, where this problem is absent.

In the present work we extend our analysis to $(3+1)$-dimensional field theory. In order to avoid problems related to the inhomogeneous distribution of the degrees of freedom, resulting from the discretization of the field on a lattice of spherical shells, we adopt an angular momentum cutoff that depends on the radius of the entangling surface in a way such that the density of degrees of freedom on the entangling surface, as well as the mean density of degrees of freedom in its interior, are independent of its radius. We extend our analysis to massive field theory, in contrast to \cite{Katsinis:2023hqn} where only massless $(1+1)$-dimensional field theory was studied. We also study cases in which only a subset of the harmonic modes lie in a squeezed state, whereas the rest lie in their ground state.

The structure of the paper is as follows: In section \ref{sec:review} we review basic facts about entanglement in squeezed states. In section \ref{sec:regularization} we analyze the discretization scheme. In section \ref{sec:results} we present our results for $(3+1)$-dimensional scalar field theory. In section \ref{sec:plateau_theory} we derive an analytic formula for the upper bound on entanglement entropy when only a subset of the modes are squeezed. Finally, in section \ref{sec:discussion} we discuss our results.

\section{Review of the method}
\label{sec:review}

An interesting property of the simple harmonic oscillator is the fact that an initial Gaussian wave function remains Gaussian at all times. Let $\omega$ be the eigenfrequency of the oscillator and $m$ its mass. Then, the most general time-dependent Gaussian solution of the simple harmonic oscillator, i.e. the wave function of a squeezed state, reads
\begin{equation}
\Psi \left( t , x \right) \sim \exp \left[ - \frac{m w \left( t \right) \left( x - x_0 \left( t \right) \right)^2}{2 \hbar} + i \frac{p_0 \left( t \right) \left( x - x_0 \left( t \right) \right)}{\hbar} \right] ,
\label{eq:squeezed}
\end{equation}
where
\begin{align}
w \left( t \right) &= \omega \frac{1 - i \sinh z \cos \left[ 2 \omega \left( t - t_0 \right) \right]}{\cosh z + \sinh z \sin \left[ 2 \omega \left( t - t_0 \right) \right]} , \label{eq:omega_squeezed} \\
x_0 \left( t \right) &= X_0 \cos \left[ \omega \left( t - t_0 \right) \right] = \left< x \right> , \label{eq:coherent_mean_x} \\
p_0 \left( t \right) &= - m \omega X_0 \sin \left[ \omega \left( t - t_0 \right) \right] = \left< p \right> . \label{eq:coherent_mean_p}
\end{align}

The squeezed states are not minimal uncertainty states. The product of the position and momentum uncertainties reads
\begin{equation}
\Delta x \Delta p = \frac{\hbar}{2} \sqrt{1 + \sinh^2 z \cos^2 \left[ 2 \omega \left( t - t_0 \right) \right]} .
\label{eq:squeezed_uncertainty_product}
\end{equation}
Therefore, these states are minimal uncertainty states at only four times per period. At any other time, the product of uncertainties may be arbitrarily high, depending of the complex parameter $z$, the so-called squeezing parameter. Only in the special case $z = 0$ they are minimal uncertainty states, the well-known coherent states. Because of the
above properties, a general squeezed state is a deep probe of the quantum characteristics of the oscillator.

In this work we calculate the reduced density matrix and the entanglement entropy in the $\left( 3 + 1 \right)$-dimensional theory of a real free scalar field, employing a direct generalization of the original method by Srednicki \cite{Katsinis:2023hqn}. This generalization concerns harmonic systems lying in a squeezed state, such as states where normal modes of the overall system are described by wave functions of the form \eqref{eq:squeezed}. 

The entanglement entropy of harmonic systems in Gaussian states, such as the squeezed states, is usually calculated via the use of the covariance matrix formalism \cite{Peschel:2002yqj,Sorkin:2012sn}. Although our approach, the direct generalization of Srednicki's wavefunction representation method, presents more technical difficulties than the covariance matrix method, we prefer it for a number of reasons: A great advantage of this method is that it provides the reduced density matrix and its spectrum. The latter contains the whole entanglement information, unlike the entanglement entropy, which is just one measure of it. More importantly, the fact that the reduced density matrix is explicitly known allows the study of the corresponding modular Hamiltonian, and, thus, the investigation of the relation between quantum entanglement and gravity in scenarios in which gravity is an emergent statistical force attributed to quantum entanglement statistics \cite{Jacobson:1995ab,VanRaamsdonk:2010pw,Jacobson:2015hqa}. This task will be the subject of future work.

Let us assume that the overall harmonic system contains $N$ degrees of freedom. Then, a general Gaussian state reads
\begin{equation}
\Psi \sim \exp \left( \left( \mathbf{x} - \mathbf{x}_0 \right)^T W \left( \mathbf{x} - \mathbf{x}_0 \right) + i \mathbf{p}_0^T \left( \mathbf{x} - \mathbf{x}_0 \right) \right) ,
\label{eq:squeezed_N}
\end{equation}
where $\mathbf{x}$ is an $N$-dimensional column matrix containing the degrees of freedom, $\mathbf{x}_0$ and $\mathbf{p}_0$ are $N$-dimensional column matrices and $W$ is an $N \times N$ symmetric matrix. Notice that in general the column matrices $\mathbf{x}_0$ and $\mathbf{p}_0$, as well as the matrix $W$, are time-dependent. More importantly, the matrix $W$ is in general complex, unlike in the ground-state case.

In general, at a given time, the above state has $N \left( N + 1 \right)$ free real parameters, which are the elements of the complex symmetric matrix $W$, and an extra $2 N$ parameters, which are the components of the vectors $\mathbf{x}_0$ and $\mathbf{p}_0$ (see \cite{Windt:2020tra} for a review on Gaussian states). In this work we restrict to states that can be written as the tensor product of a state of the form \eqref{eq:squeezed} for each of the $N$ normal modes of the system. There are two reasons for this choice: Firstly, we would like our analysis to be as simple as possible. Secondly, as the Hamiltonian of the harmonic system can also be written as the tensor product of the normal-mode Hamiltonians, such states can emerge more naturally as the time evolution of the ground state in physical setups (see for example the case of a scalar field in de Sitter space \cite{Boutivas:2023ksg,Boutivas:2023mfg}). Restricting to this kind of squeezed states reduces the number of free real parameters of the matrix $W$ from $N \left( N + 1 \right)$ to $2 N$, where the latter are simply the $N$ complex squeezing parameters of the $N$ normal modes.\footnote{Notice that the time evolution of the matrix $W$ is determined by the equation
\begin{equation}
i \frac{dW}{dt} = W^2 - \Omega^2 ,
\end{equation}
where $\Omega^2$ is the coupling matrix, i.e. the potential energy of the harmonic system reads $\frac{1}{2} \mathbf{x}^T \Omega^2 \mathbf{x}$. The solution of this equation reads
\begin{equation}
W \left( t \right) = \Omega^{1/2} \left[ I + 2 \left( e^{i \Omega t} C e^{i \Omega t} - I \right)^{-1} \right] \Omega^{1/2} ,
\end{equation}
where $C$ is a constant complex symmetric matrix determined by the initial conditions. This means that the formalism we developed is applicable to the general case as well.} There is an orthogonal matrix $O$ connecting the local coordinates $\mathbf{x}$ to the normal coordinates $\tilde{\mathbf{x}}$, i.e. $\tilde{\mathbf{x}} = O \mathbf{x}$. Then, $W = O^T \tilde{W} O$, $\mathbf{\tilde{x}}_0 = O \mathbf{x}_0 $ and $\tilde{\mathbf{p}}_{0}=O\mathbf{p}_{0}$, where $\tilde{W}$ is diagonal. The tilded matrices consist of the quantities given by equations \eqref{eq:omega_squeezed}, \eqref{eq:coherent_mean_x} and \eqref{eq:coherent_mean_p} for each normal mode of the system.

The density matrix that describes the pure state of the overall system in coordinate representation reads
\begin{multline}
\rho \left( \mathbf{x} ; \mathbf{x}^\prime \right) \sim \exp \bigg[ - \frac{1}{2} \Big( \left( \mathbf{x} - \mathbf{x}_0 \right)^T W \left( \mathbf{x} - \mathbf{x}_0 \right) \\
+ \left( \mathbf{x}^\prime - \mathbf{x}_0 \right)^T W^* \left( \mathbf{x}^\prime - \mathbf{x}_0 \right) \Big) \bigg] \exp \left[ i \mathbf{p}_{0}^T \left( \mathbf{x} - \mathbf{x}^\prime \right) \right] .
\end{multline}
Since the above density matrix is Gaussian, we may easily trace out a subset of the degrees of freedom in order to find the reduced density matrix for the rest. Without loss of generality we consider as subsystem 1 the degrees of freedom contained in the first $n$ elements of $\mathbf{x}$, and the rest as its complementary subsystem 2. We trace out the degrees of freedom of the former yielding the reduced density matrix for the latter. Adopting the block form notation
\begin{equation}
W = \left( \begin{array}{cc} A & B \\ B^T & C \end{array} \right) , \quad \mathbf{x} = \left( \begin{array}{c} \mathbf{x}_1 \\ \mathbf{x}_2 \end{array} \right) , \quad \mathbf{x}_0 = \left( \begin{array}{c} \mathbf{x}_{01} \\ \mathbf{x}_{02} \end{array} \right) , \quad \mathbf{p}_0 = \left( \begin{array}{c} \mathbf{p}_{01} \\ \mathbf{p}_{02} \end{array} \right) ,
\label{eq:many_blocks_def}
\end{equation}
the reduced density matrix reads
\begin{multline}
\rho_2 \left( \mathbf{x}_2 ; \mathbf{x}_2^\prime \right) \sim \exp \bigg[ - \frac{1}{2} \left( \left( \mathbf{x}_2 - \mathbf{x}_{02} \right)^T \gamma \left( \mathbf{x}_2 - \mathbf{x}_{02} \right) + \left( \mathbf{x}^\prime_2 - \mathbf{x}_{02} \right)^{\prime T} \gamma^* \left( \mathbf{x}^\prime_2 - \mathbf{x}_{02} \right) \right) \\
+ \left( \mathbf{x}^\prime_2 - \mathbf{x}_{02} \right)^T \beta \left( \mathbf{x}_2 - \mathbf{x}_{02} \right) + i \mathbf{p}_{02}^T \left( \mathbf{x}_2 - \mathbf{x}_2^\prime \right) \bigg] ,
\end{multline}
where
\begin{align}
\gamma &= C - \frac{1}{2} B^T \re \left( A \right)^{-1} B , \label{eq:many_gamma_def}\\
\beta &= \frac{1}{2} B^\dagger \re \left( A \right)^{-1} B . \label{eq:many_beta_def}
\end{align}
The matrix $\gamma$ is a complex symmetric matrix, whereas the matrix $\beta$ is Hermitian.

In the original calculation \cite{srednicki} the overall system lies in its ground state and the result is much simpler. Namely, the vectors $\mathbf{x}_{02}$ and $\mathbf{p}_{02}$ vanish, the matrix $\gamma$ is not just symmetric, but also real, and the matrix $\beta$ is not just Hermitian, but also real. It can be shown that all these differences, apart from the last one, do not complicate the task of calculating the eigenvalues of the reduced density matrix. The vectors $\mathbf{x}_{02}$ and $\mathbf{p}_{02}$, as well as the imaginary part of the symmetric matrix $\gamma$, do not affect the eigenvalues of $\rho_2$; they only affect the corresponding eigenstates in a trivial manner. Therefore, for the purpose of the calculation of the reduced density matrix spectrum, we can consider that $\mathbf{x}_{02}$, $\mathbf{p}_{02}$ and $\im \gamma$ vanish.

Unfortunately, the last difference is not that benign. When the matrix $\beta$ is real and symmetric, there is a combination of orthogonal transformations and rescalings of the coordinates that simultaneously diagonalizes both matrices $\gamma$ and $\beta$. This implies that the reduced density matrix can be written as the tensor product of Gaussian matrices, each one describing a single degree of freedom. These density matrices are identical to the density matrix describing a simple harmonic oscillator in a thermal state with a temperature that is determined by the corresponding eigenvalue of the matrix $\tilde{\beta} = \gamma^{-1/2} \beta \gamma^{-1/2}$. It follows that the reduced density matrix describes an effective harmonic system having $N - n$ degrees of freedom. Its normal coordinates are the coordinates that simultaneously diagonalize the matrices $\beta$ and $\gamma$. The system lies in a quasi-thermal state; each normal mode is in a thermal state, but their temperatures are different. It turns out that the eigenstates of the reduced density matrix are the tower of Fock states of this effective harmonic system, $\left| m_1 , m_2 , \ldots , m_{N - n} \right>$, where $m_i$ are non-negative integers. The corresponding eigenvalues are given by
\begin{equation}
\lambda_{\left\{ m_1 , m_2 , \ldots , m_n \right\}} = \left( 1 - \xi_1 \right) \left( 1 - \xi_2 \right) \ldots \left( 1 - \xi_n \right) \xi_1^{m_1} \xi_2^{m_2} \ldots \xi_n^{m_n} ,
\label{eq:spectrum}
\end{equation}
where the parameters $\xi_i$ are connected to the eigenvalues of $\tilde{\beta}$.

In the general case that the matrix $\beta$ is complex, there are no real coordinate transformations that simultaneously diagonalize $\gamma$ and $\beta$. However, it can be shown that the structure of the eigenstates and eigenvalues, although somehow deformed, preserves most of the properties that it has in the simpler case of \cite{srednicki}. In the remaining part of this section we present only a summary of the main facts. The complete analysis is presented in \cite{Katsinis:2023hqn}. 

Even for a complex matrix $\beta$,
there is still a tower of states. The `ground' eigenstate is again Gaussian, namely,
\begin{equation}
\left| 0 , 0 , \ldots , 0 \right> \sim \exp \left( - \frac{1}{2} \mathbf{x}_2^T \mathcal{W} \mathbf{x}_2 \right) ,
\end{equation}
where $\mathcal{W}$ is a solution to the equation
\begin{equation}
\mathcal{W} = I - \tilde{\beta}^T \left( I + \mathcal{W} \right)^{- 1} \tilde{\beta} .
\label{eq:ground_Ricatti}
\end{equation}
Then, there is again a tower of eigenstates $\left| m_1 , m_2 , \ldots , m_{N - n} \right>$ with eigenvalues given by \eqref{eq:spectrum}, where the parameters $\xi_i$ are the eigenvalues of the matrix
\begin{equation}
\Xi = \tilde{\beta}^T \left( I + \mathcal{W} \right)^{- 1} .
\label{eq:Xi}
\end{equation}
The matrix $\Xi$ is neither real nor Hermitian, yet its eigenvalues are real. If we define creation and annihilation operators that algebraically construct the tower of the eigenstates of the reduced density matrix, then these will be linear combinations of the positions and momenta, like in the simple case of \cite{srednicki}. However, the linear combination of momenta appearing in one of these operators is not the conjugate momentum of the linear combination of positions appearing in the same operator. This is the reason the reduced density matrix cannot be factored into matrices each describing a single real degree of freedom.

A complication that appears in the specification of the spectrum of the reduced density matrix is the fact that equation \eqref{eq:ground_Ricatti} is quadratic in nature, and thus it possesses many solutions. Interestingly, only one of those corresponds to normalizable eigenstates of the reduced density matrix. This is the only one that gives rise to a matrix $\Xi$ through equation \eqref{eq:Xi} whose eigenvalues are positive and smaller than one, and thus to an appropriately normalized reduced density matrix spectrum \eqref{eq:spectrum}.

The problem of selecting the right solution of equation \eqref{eq:ground_Ricatti} can be solved via the introduction of a higher-dimensional linear problem. All solutions of equation \eqref{eq:ground_Ricatti} can be constructed by the eigenvectors and eigenvalues of the $2 \left( N - n \right) \times 2 \left( N - n \right)$ matrix
\begin{equation}
M = \left( \begin{array}{cc}
2 \tilde{\beta}^{-1} & - \tilde{\beta}^{-1} \tilde{\beta}^T \\
I & O \end{array} \right) .
\end{equation}
The eigenvalues of the matrix $M$ are given by the equation
\begin{equation}
\det \left( \lambda \tilde{\beta} - 2 I + \frac{1}{\lambda} \tilde{\beta}^T \right) = 0 .
\end{equation}
Since a determinant is invariant under transposition, if follows that the $2 \left( N - n \right)$ eigenvalues of the matrix $M$ come in $N - n$ pairs of the form $\left\{ \lambda , 1 / \lambda \right\}$.
Any solution of equation \eqref{eq:ground_Ricatti} can be constructed employing $N - n$ eigenvectors and the corresponding eigenvalues of the matrix $M$. Then, the matrix $\mathcal{W}$ gives rise to a matrix $\Xi$ whose eigenvalues are exactly the inverse of the employed eigenvalues of the matrix $M$. Since the matrix $M$ has $N - n$ eigenvalues that are smaller than 1 and $N - n$ eigenvalues that are larger than 1, it follows that there is only one solution of equation \eqref{eq:ground_Ricatti} that gives rise to a normalizable spectrum of the reduced density matrix. This is the one that can be constructed by the $N - n$ eigenvectors of the matrix $M$ with eigenvalues larger than 1. It also follows that the spectrum of the correct matrix $\Xi$ is the subset of the spectrum of the matrix $M$ that contains the eigenvalues that are smaller than 1. This provides a numerical scheme for the determination of the entanglement entropy, which is given by
\begin{equation}
S_{\mathrm{EE}} = - \sum_i \left( \ln \left( 1 - \xi_i \right) + \frac{\xi_i}{1 - \xi_i} \ln \xi_i \right) ,
\label{eq:spectrum_final_SEE}
\end{equation}
as a direct consequence of equation \eqref{eq:spectrum}. The reader may consult \cite{Katsinis:2023hqn} for more details. It can be shown that this method is equivalent to the covariance matrix method \cite{Sorkin:2012sn}, see appendix E of \cite{Katsinis:2023hqn}.

\section{Discretization and regularization}
\label{sec:regularization}

\subsection{Discretization on a spherical lattice}
\label{subsec:discretization}
In order to apply the method reviewed in section \ref{sec:review} to free scalar field theory, one has to introduce a lattice discretization and reduce its dynamics to that of a quantum mechanical system with countable degrees of freedom. The simplest discretization scheme would employ a square lattice.
This simple configuration has a great advantage: the density of degrees of freedom is constant. On the other hand, choosing an entangling surface that respects the symmetries of the lattice is advantageous with respect to the complexity of the numerical calculation. If a square lattice is introduced, the natural choice of entangling surface will be a rectangular parallelepiped. Its edges and vertices  would give rise to logarithmic enhancement of some terms of the entanglement entropy. Even though the volume term cannot be enhanced, this is not the case for the area one. This fact complicates the precise determination of these terms, which are the main focus of our study.

For this reason, we insist on using a smooth spherical entangling surface. This is facilitated by the discretization of the field theory on a lattice of spherical shells following the approach of \cite{srednicki}. We start with the action of a real scalar field in flat spacetime
\begin{equation}
\mathcal{S} = \frac{1}{2} \int {d^4 x \left( \partial_\nu \phi \partial^\nu \phi - \mu^2 \phi^2 \right)} .
\end{equation}
Employing spherical coordinates this reads
\begin{equation}
\mathcal{S} = \frac{1}{2} \int {dt dr d\theta d\varphi\, r^2 \sin \theta \left( {\dot{\phi}}^2 - {\left( \partial_r \phi \right)}^2 - \frac{{\left( \partial_\theta \phi \right)}^2}{r^2} - \frac{{\left( \partial_\varphi \phi \right)}^2}{r^2 \sin^2 \theta} - \mu^2 \phi^2 \right)} .
\label{eq:scalar_action}
\end{equation}
We expand the degrees of freedom in terms of eigenfunctions of the angular momentum operator. We define the spherical harmonic moments
\begin{equation}
\phi_{\ell m} \left( t , r \right) = r \int {d\theta d\varphi \sin \theta \, Y_{\ell m} \left(\theta , \varphi \right) \phi \left(t , r , \theta , \varphi \right) } ,
\label{eq:spherical_harmonic_moments}
\end{equation}
where $Y_{\ell m} \left(\theta , \varphi \right)$ are the real spherical harmonics. The scalar field is a linear combination of the spherical harmonic moments, namely,
\begin{equation}
\phi = \sum_{\ell , m} {\frac{\phi_{\ell m} Y_{\ell m}}{r}} .
\label{eq:scalar_field_decomposition}
\end{equation}
Substituting \eqref{eq:scalar_field_decomposition} in \eqref{eq:scalar_action} yields
\begin{equation}
\mathcal{S} = \frac{1}{2} \sum_{\ell , m} \int {dt dr \left[ \dot{\phi}_{\ell m}^2 - r^2 {\left[ \partial_r \left( \frac{\phi_{\ell m}}{r} \right) \right]}^2 - \left( \frac{\ell \left( \ell + 1 \right)}{r^2} + \mu^2 \right) \phi_{\ell m}^2 \right]} .
\label{eq:action_moments}
\end{equation}

The only continuous variable left is the radial coordinate $r$. Following \cite{srednicki}, we discretize the radial coordinate introducing a lattice of spherical shells with radii $r_i = i a$, where $i = 1 , 2 , \ldots , N$. The radial distance $a$ between adjacent spherical shells imposes an UV cutoff, while the overall size of the lattice $N a$ imposes an IR cutoff. We use the discretization scheme:
\begin{align}
r &\to ja\\
\phi_{\ell m} \left( r \right) &\to \phi_{\ell m , j} , \\
{\left. \frac{\partial \phi_{\ell m} \left( r \right)}{\partial r} \right|}_{r = j a} &\to \frac{\phi_{\ell m , j + 1} - \phi_{\ell m , j}}{a} , \\
\int_0^{\left( {N + 1} \right) a} {dx}  &\to a\sum\limits_{j = 1}^N {} .
\label{eq:discetization}
\end{align}
The discretized action reads
\begin{equation}
\mathcal{S} =\int dt \sum\limits_{\ell , m} {L_{\ell m}} ,
\label{eq:action_discetized}
\end{equation}
where
\begin{equation}
L_{\ell m} = \frac{a}{2} \sum_{j = 1}^N {\Bigg[ \dot{\phi}_{\ell m , j}^2 - \frac{{\left( {j + \frac{1}{2}} \right)}^2}{a^2} {\left( \frac{\phi_{\ell m , j + 1}}{j + 1} - \frac{\phi_{\ell m , j}}{j} \right)}^2} \\
{- \left( \mu^2 + \frac{\ell \left( \ell + 1 \right)}{j^2 a^2} \right) {\phi_{\ell m , j}}^2 \Bigg]} .
\label{eq:sector_Lagrangian}
\end{equation}
This action describes an infinite, but countable set of harmonic systems. Each one is identified by the pair $\left\{ \ell , m \right\}$ and has dynamics described by the Lagrangian \eqref{eq:sector_Lagrangian}. These systems do not interact with each other and, furthermore, their dynamics does not depend on $m$. Concerning entanglement entropy, the above implies a $2 \ell + 1$ degeneracy for the contribution of each $\ell$-sector.

We define the canonical momenta
\begin{equation}
\pi_{\ell m , j} = a \dot{\phi}_{\ell m , j} .
\label{eq:canonical_momanta}
\end{equation}
Then, the Hamiltonian that corresponds to the Lagrangian \eqref{eq:sector_Lagrangian} reads
\begin{equation}
H_{\ell m} = \frac{1}{2 a} \sum\limits_{j = 1}^N {\left[ \pi_{\ell m , j}^2 + {\left( {j + \frac{1}{2}} \right)}^2 {\left( \frac{\phi_{\ell m , j + 1}}{j + 1} - \frac{\phi_{\ell m , j}}{j} \right)}^2 + \left( \mu^2a^2 + \frac{\ell \left( \ell + 1 \right)}{j^2} \right) {\phi_{\ell m , j}}^2 \right]} .
\label{eq:sector_Hamiltonian}
\end{equation}
Thus, due to the spherical symmetry of the setup, the Hamiltonian becomes an infinite sum of Hamiltonians, one for each $\ell$-sector. We define as subsystem 1 the degrees of freedom contained in the spherical shells numbered with indices $i \leq n$, whereas the subsystem 2 is the complementary subsystem. This is equivalent to considering a spherical entangling surface with radius $R \simeq \left( n + \frac{1}{2} \right) a$.

We should point out that, even though the discretization scheme is enforced by the symmetry of the problem, it is rather peculiar in the context of traditional effective field theory. There is no cutoff of the angular degrees of freedom. So, essentially we are describing a system as perceived by an observer who can probe modes of arbitrarily low wavelength in the angular directions, but up to a finite wavelength in the radial direction. Interestingly enough, this scheme works as a method of regularization without imposing a cutoff on the angular momentum.\footnote{This is true in $2 + 1$ and $3 + 1$ dimensions. In $4 + 1$ or more dimensions an angular cutoff is required.} This is due to the fact that the angular momentum acts as an effective position-dependent mass, which localizes the normal modes of \eqref{eq:sector_Hamiltonian} and suppresses entanglement. It turns out that the series 
\begin{equation}
S_{\mathrm{EE}} = \sum_{\ell=0}^{\infty} \left( 2 \ell + 1 \right) S_{\mathrm{EE} , \ell}
\label{eq:entropy_series}
\end{equation}
is convergent. At the practical level, however, when numerical calculations are performed an angular cutoff is imposed.

\subsection{The $\ell_{\max}\sim n$ regularization}
\label{subsec:reg_lmax}
Without an angular cutoff each spherical shell of the lattice contributes an infinite number of degrees of freedom. In the continuous limit, i.e. when the number of shells goes to infinity and the lattice spacing to zero, the degrees of freedom are distributed homogeneously. However, this is not the case when the lattice spacing is finite. Having an inhomogeneous distribution of degrees of freedom affects the scaling of entanglement entropy. Imposing a cutoff $\ell_{\textrm{max}}$ to the angular momentum implies that in each spherical shell there are $\left( \ell_{\textrm{max}} + 1 \right)^2$ degrees of freedom, independently of the radius of the shell. In an obvious manner, the density of the degrees of freedom is not constant in space, but it is a decreasing function of the radial coordinate.

As long as one studies the states of the theory, like the ground state, in which the entanglement entropy is dominated by an area-law term, things are relatively simple. The area-law term is a local term. It emerges due to entanglement between neighbouring degrees of freedom that are separated by the entangling surface. Therefore, in order to correctly study the scaling properties of an area-law term, one should ensure that the density of the degrees of freedom on the entangling surface does not change as we change its radius. This is achieved by considering an $\ell_{\textrm{max}}$ that depends on the radius of the entangling surface, as was done in \cite{Katsinis:2017qzh}, namely
\begin{equation}
\ell_{\textrm{max}} \simeq c n .
\label{eq:lmax}
\end{equation}
For any finite value of $c$, the entanglement entropy is dominated by an area-law term, with a smaller coefficient than in the limit $c \rightarrow \infty$ taken in \cite{srednicki}. This difference is not unexpected, as this coefficient is scheme dependent.

The above is sufficient for the study of entanglement entropy when this is dominated by an area-law term. In the case at hand, we expect a volume term to develop as well \cite{Katsinis:2023hqn}. Thus, it is necessary that both the volume and the area terms are not disturbed by the inhomogeneous distribution of the degrees of freedom in our discretization scheme. Fortunately, the scheme \eqref{eq:lmax} preserves both the density of degrees of freedom on the entangling surface and the mean density of degrees of freedom in its interior. In particular, these read
\begin{equation}
\rho_{\textrm{surface}} = \frac{\left( \ell_{\textrm{max}} + 1 \right)^2}{4 \pi R^2} \simeq \frac{c^2}{4 \pi a^2},\qquad
\bar{\rho}_{\textrm{volume}} = \frac{n \left( \ell_{\textrm{max}} + 1 \right)^2}{\frac{4}{3} \pi R^3} \simeq \frac{3 c^2}{4 \pi a^3} .
\end{equation}
The scheme works in any number of dimensions. 
Demanding  $\rho_{\textrm{surface}} = {1}/{a^2}$, so that the density of the degrees of freedom is isotropic in the region of the entangling surface, implies that
\begin{equation}
c = \sqrt{4 \pi} .
\label{eq:lmax_sim_n_coeff}
\end{equation}
In what follows, we shall call this approach the ``$\ell_{\max} \sim n$'' regularization and use the convention \eqref{eq:lmax_sim_n_coeff}. 
\pagebreak
\section{Numerical results}
\label{sec:results}

In this section we present results that demonstrate the effect of squeezing on entanglement in $( 3 + 1 )$-dimensional free scalar field theory. In all cases we use a lattice having $N = 60$ spherical shells as in the original calculation by Srednicki \cite{srednicki} and set the lattice spacing $a$ equal to 1, i.e. we measure lengths in units of the UV cutoff. As we discussed in section \ref{sec:regularization}, we divide the degrees of freedom using a spherical entangling surface with radius $R \simeq \left( n + \frac{1}{2} \right) a$.

An important question we want to address is whether there is a contribution to the entanglement entropy that is proportional to the volume of the smaller of the two subsystems, in line with Page's argument \cite{Page:1993df} and with results in $(1 + 1)$-dimensional field theory \cite{Katsinis:2023hqn}. Naturally, in infinite three-dimensional space, the smaller subsystem is always the one inside the spherical entangling surface. However, when we discretize the field theory on a lattice, this is not manifest. For this reason, we restrict ourselves to $n \leq 30 = {N}/{2}$.

Previous works on entanglement entropy in field theory, e.g. \cite{srednicki, Katsinis:2017qzh}, study the scaling properties of the entropy with the size of the considered subsystem, usually a spherical region, similarly to this study. Most calculations are performed in the ground state, which obviously has both translational and rotational symmetry. Having the above in mind, in appendix \ref{app:state} we show that, for a state with the squeezing parameter of each mode given by function $z \left( \omega \right)$ of the eigenfrequency of the normal mode, the \emph{mean over time} of any local observable preserves the symmetries of the Hamiltonian, although the symmetries are broken at any given time. For this reason, instead of studying the entanglement entropy at a given instant, we study the mean entanglement entropy over time. In order to do so, throughout the numerical calculations we sample over 100 random time instances.

It must be also noted that, when we discretize the field theory on a finite lattice of spherical shells, translational invariance is explicitly broken. However, in the limit that the lattice becomes very large, the translational invariance of the mean over time of local observables is recovered in the same way as the translational invariance of the dynamics of the theory. Therefore, our specific choice of state ensures that the discretized system has a meaningful field theory limit.

Although the available computational power has obviously increased since the era of \cite{srednicki}, we kept the same lattice size because the calculations in the case of squeezed states require a much larger number of significant digits. In some of our calculations 3000 significant digits were required. Furthermore, since the entanglement entropy is time dependent and the desired outcome is its mean value, each calculation has to be repeated for several random times. Finally, our calculations are performed for various values of the squeezing parameter and the field mass in order to investigate the dependence of entanglement entropy on these parameters, further multiplying the number of required numerical runs.

\subsection{Uniform squeezing of all modes---massless field theory}
\label{subsec:results_massless}

We consider free massless scalar field theory in $3 + 1$ dimensions, when all modes lie in a squeezed state with the same squeezing parameter $z$. Deeper understanding of the behaviour of entanglement entropy is facilitated by the study of the contribution of each angular momentum sector separately, as described by equation \eqref{eq:entropy_series}. The dynamics of each sector is governed by the Hamiltonian \eqref{eq:sector_Hamiltonian}, which can be considered as the discretized version of a $(1+1)$-dimensional field theory with a position-dependent mass term. Therefore, we may understand the behaviour of these contributions to the entanglement entropy by comparing them to the free massless $(1+1)$-dimensional free field theory that has already been studied in \cite{Katsinis:2023hqn}.

The entanglement entropy in free massless $(1+1)$-dimensional theory at its ground state is dominated by a logarithmic universal term \cite{Calabrese:2004eu,Calabrese:2009qy}. However, when all the modes are squeezed, even with a small squeezing parameter $z$, a volume term appears. This term is proportional to the number of the degrees of freedom of the smaller subsystem, i.e. $S_{\mathrm{volume}} \sim \min \left( n , N - n \right)$. This volume-law behaviour is expected to be common for any harmonic system at an arbitrary quantum state following Page's argument \cite{Page:1993df}.\footnote{A typical example of a state with entanglement entropy following a volume law is a thermal state, see for instance \cite{Katsinis:2019vhk,Katsinis:2019lis}. A volume dependence has also been observed in field theories on time-dependent backgrounds \cite{Boutivas:2023mfg}, theories with broken Lorentz invariance	\cite{mozaffar1}, and multi-field theories with ``field-space entanglement'' \cite{mollbashi,mozaffar2}.} In \cite{Katsinis:2023hqn} a large-squeezing expansion was performed for an arbitrary harmonic system. It turns out that, at the limit of large $z$, the entanglement entropy is time-independent and approximately equal to
\begin{equation}
S \simeq z \min \left( n , N - n \right) .
\label{eq:large_sq_one_d}
\end{equation}
In the same work it was shown that for $z > 25$, this formula is a good approximation for the entanglement entropy in $(1+1)$-dimensional massless field theory.

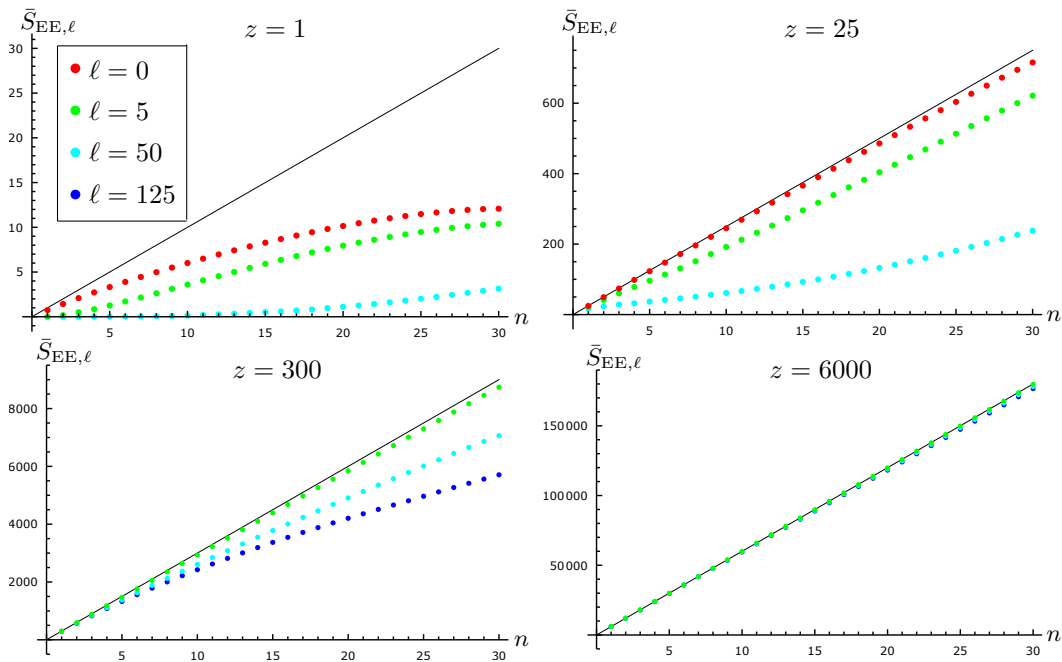
\begin{figure}[ht]
\centering
\begin{picture}(92,59)
\put(0,0){\includegraphics[angle=0,width=0.9\textwidth]{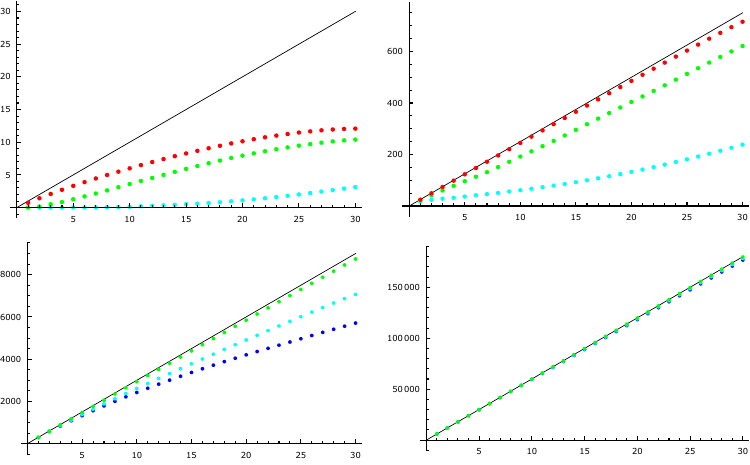}}
\put(4,39){\includegraphics[angle=0,width=0.1\textwidth]{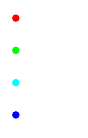}}
\put(4.25,39.25){\line(1,0){11.25}}
\put(4.25,39.25){\line(0,1){15.375}}
\put(4.25,54.625){\line(1,0){11.25}}
\put(15.5,39.375){\line(0,1){15.375}}
\put(44,2){{\footnotesize $n$}}
\put(90.375,2.375){{\footnotesize $n$}}
\put(44,30.25){{\footnotesize $n$}}
\put(90.375,30.5){{\footnotesize $n$}}
\put(2.375,27.125){{\footnotesize $\bar{S}_{\mathrm{EE},\ell}$}}
\put(50.25,26.625){{\footnotesize $\bar{S}_{\mathrm{EE},\ell}$}}
\put(1,56.125){{\footnotesize $\bar{S}_{\mathrm{EE},\ell}$}}
\put(48.125,56){{\footnotesize $\bar{S}_{\mathrm{EE},\ell}$}}
\put(19.5,25.5){{\small $z=300$}}
\put(66.25,25.5){{\small $z=6000$}}
\put(20.5,55.375){{\small$z=1$}}
\put(67.5,55.375){{\small$z=25$}}
\put(7,40.875){{\small$\ell = 125$}}
\put(7,44.5){{\small$\ell = 50$}}
\put(7,48.125){{\small$\ell = 5$}}
\put(7,51.75){{\small$\ell = 0$}}
\end{picture}
\caption{The contribution of a single angular momentum sector to the mean entanglement entropy as a function of $n$. The continuous line is the large-squeezing approximation \cite{Katsinis:2023hqn}, namely $\bar{S}_{\mathrm{EE}} = z n$.}
\label{fig:massless_sectors_smallz}
\end{figure}

Figure \ref{fig:massless_sectors_smallz} depicts the contributions of various angular momentum sectors to the entanglement entropy.
The $\ell = 0$ sector is actually identical to the $(1+1)$-dimensional free massless  theory. The figure displays some characteristic features:
\begin{itemize}
\item The contributions of the higher-angular momentum sectors are smaller than that of the vanishing angular momentum sector.
\item The larger the angular momentum, the larger is the deviation from the $\ell = 0$ sector.
\item All contributions are increasing functions of the squeezing parameter $z$.
\item For $z\sim 25$ the $\ell = 0$ sector is well-approximated by the large-squeezing expansion. The other sectors converge to the large-squeezing approximation for larger values of $z$.
\end{itemize}

The Hamiltonians of the sectors with $\ell > 0$ differ from that of the $\ell = 0$ sector in the diagonal terms, i.e. the self-couplings of the degrees of freedom, which depend on position. In the limit that the self-couplings of the local oscillators become much larger than the couplings to their neighbours, which is the limit of large $\ell$, the modes become effectively localized;\footnote{It is crucial that the self-couplings that are induced by the angular momentum are position-dependent. A mass term would generate position-independent self-couplings, i.e. a contribution to the couplings matrix that is proportional to the identity. This does not localize the normal modes. We will return to this issue in section \ref{subsec:results_mass}, where we study massive field theory.} each mode affects a single local degree of freedom and the local oscillators behave as if they were decoupled. It follows that it is impossible to enforce entanglement between different sites of the lattice via these modes, independently of whether they are squeezed or not.\footnote{This is the very essence of the inverse mass expansion for the entanglement entropy at the ground state performed in \cite{Katsinis:2017qzh}, where a perturbative expansion is developed around this limit.}

Nevertheless, at large squeezing, all harmonic systems exhibit the same behaviour. As shown in figure \ref{fig:massless_sectors_smallz}, in this limit the contributions of all angular momentum sectors are well-approximated by \eqref{eq:large_sq_one_d}. The critical value of $z$ for which the behaviour of an angular momentum sector approaches that of the large-squeezing expansion is an increasing function of $\ell$. In this work, because of the specific discretization that we use, we deal with sectors with angular momentum up to $\ell_{\max} \simeq 125$. All these sectors are dominated by the volume term of the large-squeezing expansion for squeezing parameters of order 1000.

\begin{figure}[p]
\centering
\begin{picture}(92,59)
\put(0,0){\includegraphics[angle=0,width=0.9\textwidth]{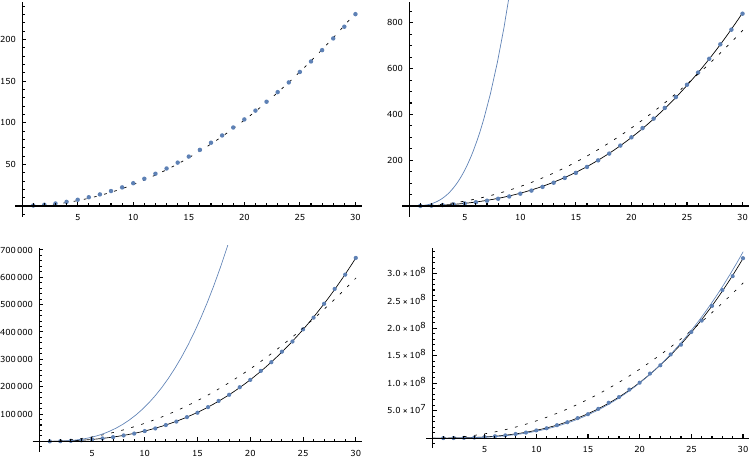}}
\put(44,2.375){{\footnotesize $n$}}
\put(90.5,2.75){{\footnotesize $n$}}
\put(44,30.625){{\footnotesize $n$}}
\put(90.375,30.625){{\footnotesize $n$}}
\put(3.75,26.375){{\footnotesize $\bar{S}_{\mathrm{EE}}$}}
\put(51,26.375){{\footnotesize $\bar{S}_{\mathrm{EE}}$}}
\put(1.75,55.875){{\footnotesize $\bar{S}_{\mathrm{EE}}$}}
\put(48.125,55.875){{\footnotesize $\bar{S}_{\mathrm{EE}}$}}
\put(20.25,25.875){{\small $z=10$}}
\put(66.5,25.875){{\small $z=1000$}}
\put(20.75,55.125){{\small$z=0$}}
\put(67.5,55.125){{\small$z=0.1$}}
\end{picture}
\caption{The mean entanglement entropy as a function of $n$. The blue continuous line is the large-squeezing approximation \eqref{eq:large_sq_lmax_reg}. The dashed black line is a fit to the numerical data by a single area term (equation \eqref{eq:fit_vacuum}), whereas the continuous black one is a fit to the numerical data by an area and a volume term (equation \eqref{eq:fit_squeeze}).}
\label{fig:massless_ell_reg}
\begin{picture}(93,61)
\put(0,0){\includegraphics[angle=0,width=0.9\textwidth]{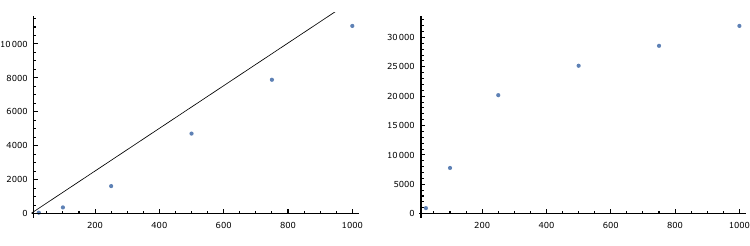}}
\put(0,29){\includegraphics[angle=0,width=0.9\textwidth]{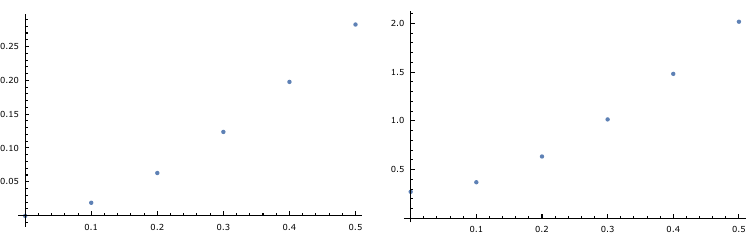}}
\put(43.5,3.375){{\footnotesize $z$}}
\put(90,3.375){{\footnotesize $z$}}
\put(44,32.25){{\footnotesize $z$}}
\put(90,31.875){{\footnotesize $z$}}
\put(3.25,28.25){{\footnotesize $a_3$}}
\put(49.75,28.25){{\footnotesize $a_2$}}
\put(2.125,57.5){{\footnotesize $a_3$}}
\put(48.375,57.75){{\footnotesize $a_2$}}
\end{picture}
\caption{The coefficients $a_2$ and $a_3$ as functions of $z$. The continuous line in the bottom left panel is the large-squeezing approximation, namely $a_3 \simeq 4 \pi z$.}
\label{fig:massless_ell_reg_coeff}
\end{figure}

Employing equation \eqref{eq:entropy_series}, we sum the contributions of all relevant angular momentum sectors and calculate the total entanglement entropy in $(3+1)$-dimensional field theory. It is well-known \cite{srednicki} that in the ground state ($z = 0$) the entanglement entropy is dominated by an area-law term, i.e. it is well-fitted by a curve of the form
\begin{equation}
S_{\mathrm{EE}} \simeq a_2 n^2 .
\label{eq:fit_vacuum}
\end{equation}
This is visible in the top-left panel of figure \ref{fig:massless_ell_reg}, where the dashed black lines are fits to the numerical data by a single area-law term.

As we discussed in the introduction, this is a fact that contradicts Page's argument \cite{Page:1993df}, which states that the entanglement entropy in an arbitrary quantum state of the system scales with the number of the degrees of freedom of the smaller subsystem, i.e. its volume. Indeed, when squeezing is turned on, as is visible in figure \ref{fig:massless_ell_reg}, the fit \eqref{eq:fit_vacuum} is not good, even for relatively small values of the squeezing parameter. On the contrary, if a volume term is included in the fitting curve, so that
\begin{equation}
\bar{S}_{\mathrm{EE}} \simeq a_3 n^3 + a_2 n^2 ,
\label{eq:fit_squeeze}
\end{equation}
the curve fits almost perfectly the numerical data.

If the squeezing is large enough, so that all angular momentum sectors are well-approximated by the large-squeezing expansion, the entropy is dominated by a volume term given by a simple analytic expression. As we stated above, this is achieved at squeezing parameters of order 1000. Recalling that in our regularization each spherical shell contains $\ell_{\max}^2 = 4 \pi n^2$ degrees of freedom, the number of  degrees of freedom in the interior of the entangling sphere is approximately $4 \pi n^3$. We restrict ourselves to the cases where $n \leq N/2$, so that the interior of the entangling surface contains fewer degrees of freedom than the exterior. Then, the large-squeezing approximation reads
\begin{equation}
\bar{S}_{\mathrm{EE}} \simeq 4 \pi z n^3 ,
\label{eq:large_sq_lmax_reg}
\end{equation}
which is indeed a good approximation of the entanglement entropy for $z = 1000$, as shown in the bottom-right panel of figure \ref{fig:massless_ell_reg}.

The dependence of the coefficients $a_2$ and $a_3$ is shown in figure \ref{fig:massless_ell_reg_coeff}. 
Both coefficients are increasing functions of $z$. As $z \to 0$ the coefficient $a_3$ of the volume term tends to zero as we expect, since the volume term is absent in the case of the ground state. For small values of $z$ this coefficient is quadratic in $z$, as expected by the small-squeezing expansion presented in \cite{Katsinis:2023hqn}. For large squeezing parameters it becomes a linear function of $z$ in line with the large-squeezing expansion, which suggests that in this limit $a_3 \simeq 4 \pi z$. As $z \to 0$ the coefficient $a_2$ of the area term tends to the finite value $a_2 \simeq 0.27$ coinciding with the value found both numerically and via a perturbative expansion for the ground state in the $\ell_{\max}\sim n$ regularization in \cite{Katsinis:2017qzh}. 

\subsection{Uniform squeezing of all modes---massive field theory}
\label{subsec:results_mass}

In this section we study entanglement in massive scalar field theory in a squeezed state. It is known that the mass of the field generally suppresses entanglement in the ground state. We have already seen that position-dependent self-couplings suppress entanglement due to the localization of the normal modes. This is a mechanism that we investigated in section \ref{subsec:results_massless}. However, the field mass term corresponds to a position-independent self-coupling. As a result, it does not alter the shape of the normal modes, but only their frequencies. Although mass suppresses entanglement in the ground state, it is not obvious whether it does so in squeezed states of the system.
\begin{figure}[p]
\centering
\begin{picture}(92,64)
\put(0,5){\includegraphics[angle=0,width=0.9\textwidth]{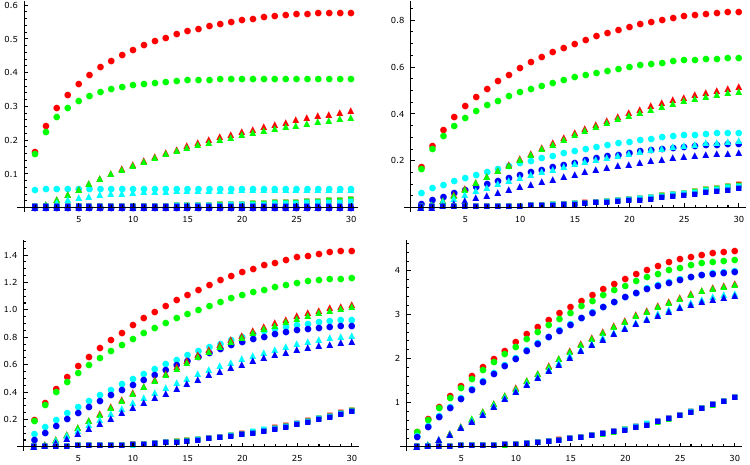}}
\put(54,0.5){\includegraphics[angle=0,width=0.3\textwidth]{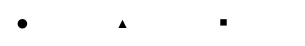}}
\put(2.5,0){\includegraphics[angle=0,width=0.4\textwidth]{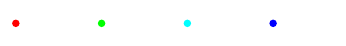}}
\put(43.625,6.625){{\footnotesize $n$}}
\put(89.5,6.625){{\footnotesize $n$}}
\put(43.5,35.375){{\footnotesize $n$}}
\put(89.875,35.375){{\footnotesize $n$}}
\put(2,32.375){{\footnotesize $\bar{S}_{\mathrm{EE},\ell}$}}
\put(48,32.375){{\footnotesize $\bar{S}_{\mathrm{EE},\ell}$}}
\put(2,61){{\footnotesize $\bar{S}_{\mathrm{EE},\ell}$}}
\put(48.5,61){{\footnotesize $\bar{S}_{\mathrm{EE},\ell}$}}
\put(19.5,30.875){{\small $z=0.2$}}
\put(67,30.875){{\small $z=0.5$}}
\put(20.5,60.125){{\small$z=0$}}
\put(67,60.125){{\small$z=0.1$}}
\put(5,2){{\small$\mu=0$}}
\put(15,2){{\small$\mu=\frac{1}{10}$}}
\put(25,2){{\small$\mu=1$}}
\put(35,2){{\small$\mu=5$}}
\put(57,2){{\small$\ell = 0$}}
\put(67,2){{\small$\ell = 5$}}
\put(77,2){{\small$\ell = 50$}}
\end{picture}
\caption{The contribution of a single angular momentum sector to the mean entanglement entropy as a function of $n$ for various field masses.}
\label{fig:massive_sectors}
\begin{picture}(92,64)
\put(0,5){\includegraphics[angle=0,width=0.9\textwidth]{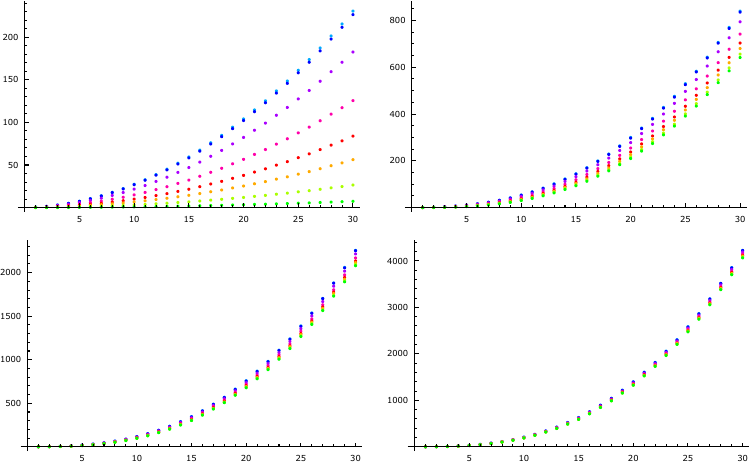}}
\put(6,0){\includegraphics[angle=0,width=0.8\textwidth]{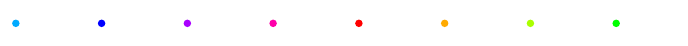}}
\put(44,6.625){{\footnotesize $n$}}
\put(90.5,6.625){{\footnotesize $n$}}
\put(43.75,35.375){{\footnotesize $n$}}
\put(90.25,35.375){{\footnotesize $n$}}
\put(2.5,32.5){{\footnotesize $\bar{S}_{\mathrm{EE}}$}}
\put(49,32.5){{\footnotesize $\bar{S}_{\mathrm{EE}}$}}
\put(2.25,61.25){{\footnotesize $\bar{S}_{\mathrm{EE}}$}}
\put(48.75,61.25){{\footnotesize $\bar{S}_{\mathrm{EE}}$}}
\put(20,30.875){{\small $z=0.2$}}
\put(67,30.875){{\small $z=0.3$}}
\put(20.75,60.125){{\small$z=0$}}
\put(67,60.125){{\small$z=0.1$}}
\put(8.5,2){{\small$\mu=0$}}
\put(18.5,2){{\small$\mu=\frac{1}{10}$}}
\put(28.5,2){{\small$\mu=\frac{1}{2}$}}
\put(38.5,2){{\small$\mu=1$}}
\put(48.5,2){{\small$\mu=\frac{3}{2}$}}
\put(58.5,2){{\small$\mu=2$}}
\put(68.5,2){{\small$\mu=3$}}
\put(78.5,2){{\small$\mu=5$}}
\end{picture}
\caption{The mean entanglement entropy as a function of $n$ for various field masses.}
\label{fig:massive_ell_reg}
\end{figure}

In figure \ref{fig:massive_sectors}, 
the contribution of a single angular momentum sector to the mean entanglement entropy is depicted for various values of the field mass. For very small values of $z$, the mass appears to reduce entanglement, as in the ground state of the system. However, when $z$ increases, the entanglement entropy becomes insensitive to the value of the mass. This behaviour is consistent with the form of the large-squeezing expansion in arbitrary harmonic systems \cite{Katsinis:2023hqn}. The expansion parameter is the squeezing parameter $z$, which is a pure number and cannot be compared to any dimensionful parameter, such as the field mass. Therefore, the same large-squeezing limit applies to all harmonic systems, independently of whether they are massive or not.

We calculate the mean entanglement entropy in $( 3 + 1 )$-dimensional massive field theory using the $\ell_{\max} \sim n$ regularization via the addition of the contributions of all relevant angular momentum sectors. The results are depicted in figure \ref{fig:massive_ell_reg}.
\begin{figure}[t]
\centering
\begin{picture}(93,36.5)
\put(0,5){\includegraphics[angle=0,width=0.9\textwidth]{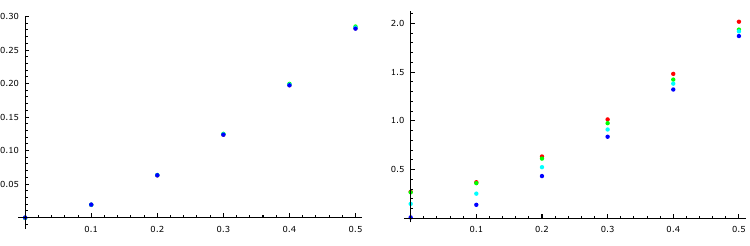}}
\put(27.5,0){\includegraphics[angle=0,width=0.4\textwidth]{massive_ell_sectors_small_z_legend_colors.pdf}}
\put(43.875,8){{\footnotesize $z$}}
\put(90,8){{\footnotesize $z$}}
\put(2.25,33.375){{\footnotesize $a_3$}}
\put(48.5,33.675){{\footnotesize $a_2$}}
\put(30,2){{\small$\mu=0$}}
\put(40,2){{\small$\mu=\frac{1}{10}$}}
\put(50,2){{\small$\mu=1$}}
\put(60,2){{\small$\mu=5$}}
\end{picture}
\caption{The coefficients $a_2$ and $a_3$ as function of $z$ for various field masses.}
\label{fig:massive_ell_reg_coeff}
\end{figure}
The numerical data for all masses are very well-fitted by the sum of an area and a volume term. The volume term vanishes at $z = 0$ for all masses. As $z$ increases, the entanglement entropy becomes insensitive to the value of the mass. Actually, this behaviour becomes apparent at relatively small values of $z$.

The above fact is also evident in the coefficients of the area and volume terms, which are depicted in figure \ref{fig:massive_ell_reg_coeff}. Notice that the volume term is essentially independent of the value of the mass for all values of $z$.

\subsection{Squeezing of a subset of modes}
\label{subsec:results_plateau}

In this section we focus on massless free scalar field theory. The Hamiltonian governing the dynamics of each $\ell$-sector is given by equation \eqref{eq:sector_Hamiltonian} with $\mu=0$. Initially we consider the $\ell = 0$ sector in a state in which the $N_{\textrm{sq}} < N$ lowest frequency modes lie in a squeezed state with the same parameter $z$, while the remaining of the modes lie in their ground state.

\begin{figure}[t]
\centering
\begin{picture}(92,59.5)
\put(0,0){\includegraphics[angle=0,width=0.9\textwidth]{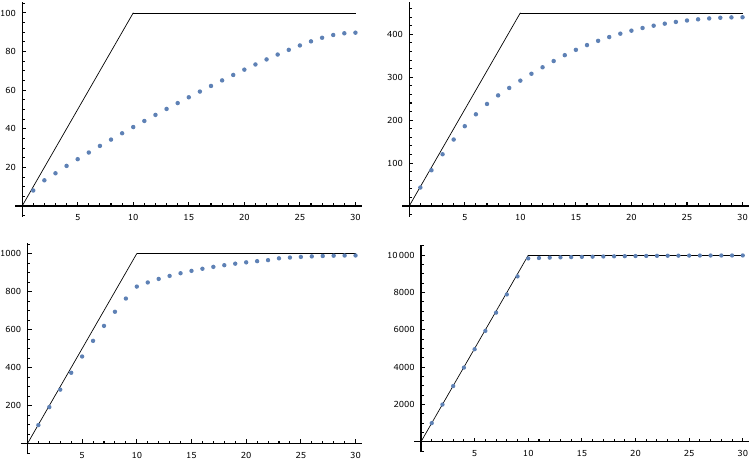}}
\put(44,2){{\footnotesize $n$}}
\put(90.375,2.25){{\footnotesize $n$}}
\put(44,30.5){{\footnotesize $n$}}
\put(90.375,30.5){{\footnotesize $n$}}
\put(2.375,27.125){{\footnotesize $\bar{S}_{\mathrm{EE}}$}}
\put(49.125,27){{\footnotesize $\bar{S}_{\mathrm{EE}}$}}
\put(1.75,56){{\footnotesize $\bar{S}_{\mathrm{EE}}$}}
\put(48.25,56){{\footnotesize $\bar{S}_{\mathrm{EE}}$}}
\put(20,26.5){{\small $z=100$}}
\put(67.75,26.5){{\small $z=1000$}}
\put(20.5,56.375){{\small$z=10$}}
\put(68.25,56.375){{\small$z=45$}}
\multiput(16.375,3.75)(0,2){11}{\line(0,1){1}}
\multiput(63.5,3.75)(0,2){11}{\line(0,1){1}}
\multiput(16,32.25)(0,2){11}{\line(0,1){1}}
\multiput(62.5,32.25)(0,2){11}{\line(0,1){1}}
\multiput(4.75,25.25)(2,0){6}{\line(1,0){1}}
\multiput(4.25,54.25)(2,0){6}{\line(1,0){1}}
\multiput(51.875,25.125)(2,0){6}{\line(1,0){1}}
\multiput(50.375,54.25)(2,0){6}{\line(1,0){1}}
\end{picture}
\caption{The mean entanglement entropy as a function of $n$ for $N = 60$, $N_{\textrm{sq}} = 20$ and various $z$. The continuous line is the approximation given by equation \eqref{eq:plateau}.}
\label{fig:plateau_massless_1d}
\end{figure}
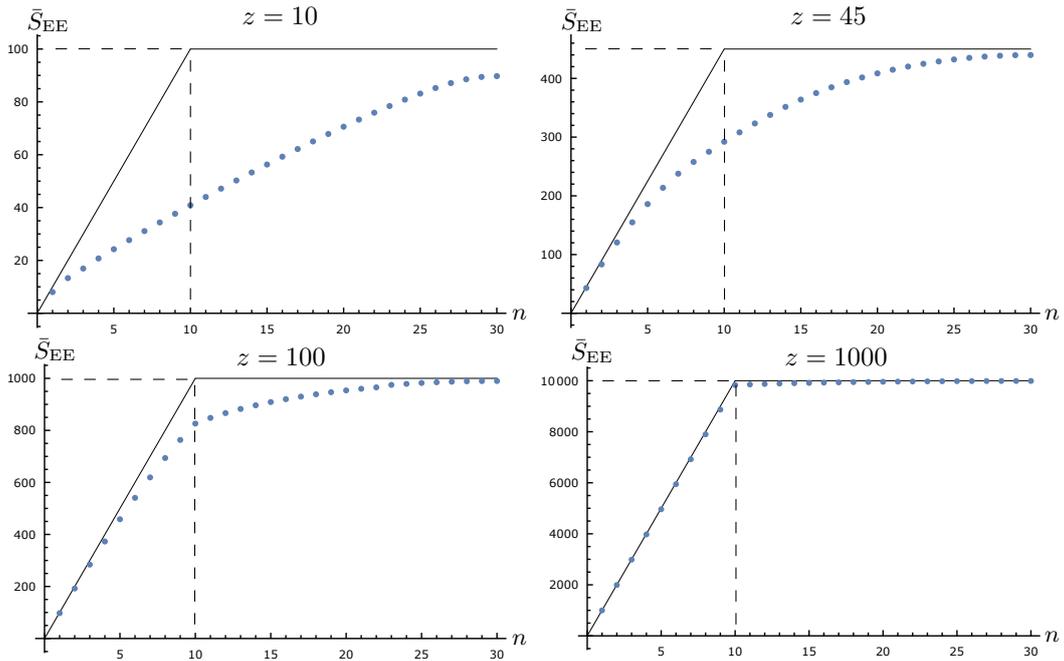

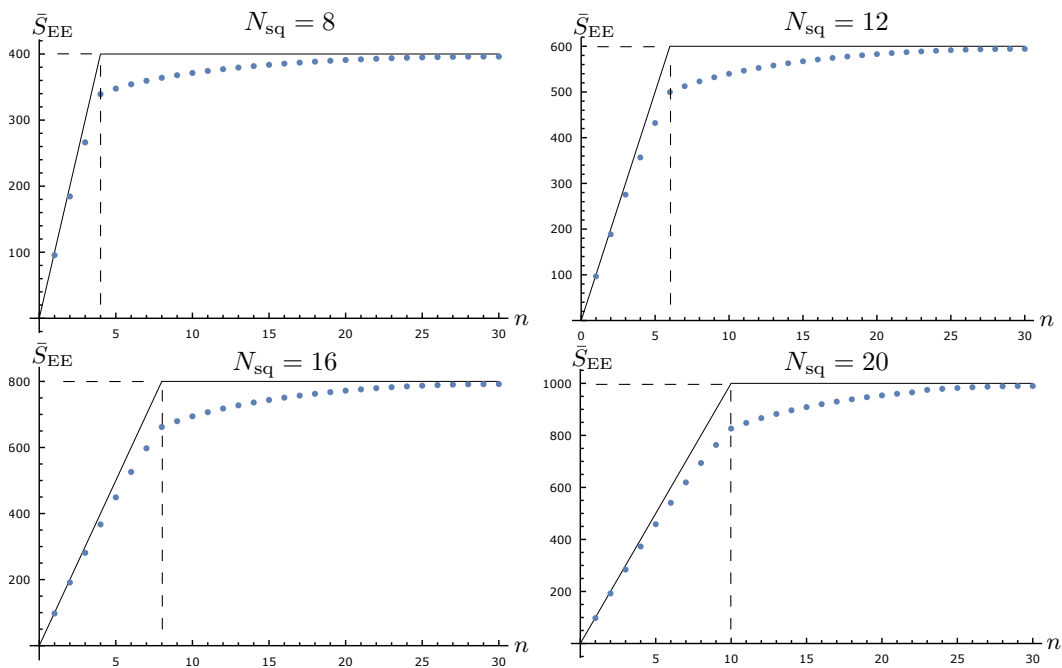
\begin{figure}[ht]
\centering
\begin{picture}(92,59.5)
\put(0,0){\includegraphics[angle=0,width=0.9\textwidth]{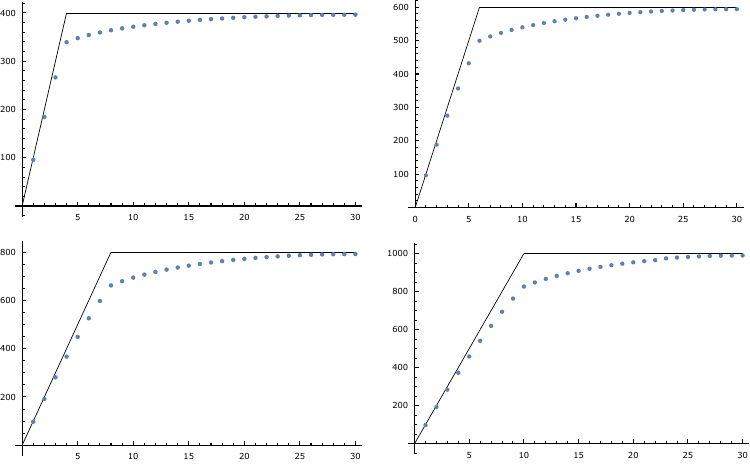}}
\put(44,1.75){{\footnotesize $n$}}
\put(90.375,2){{\footnotesize $n$}}
\put(44,30.5){{\footnotesize $n$}}
\put(89.75,30.375){{\footnotesize $n$}}
\put(2,27.125){{\footnotesize $\bar{S}_{\mathrm{EE}}$}}
\put(49,26.875){{\footnotesize $\bar{S}_{\mathrm{EE}}$}}
\put(2,56){{\footnotesize $\bar{S}_{\mathrm{EE}}$}}
\put(49.125,56){{\footnotesize $\bar{S}_{\mathrm{EE}}$}}
\put(19.5,26.5){{\small $N_{\textrm{sq}}=16$}}
\put(67.5,26.5){{\small $N_{\textrm{sq}}=20$}}
\put(20.5,56.375){{\small$N_{\textrm{sq}}=8$}}
\put(67.5,56.375){{\small$N_{\textrm{sq}}=12$}}
\multiput(13.375,3.75)(0,2){11}{\line(0,1){1}}
\multiput(62.875,3.75)(0,2){11}{\line(0,1){1}}
\multiput(8,32.25)(0,2){11}{\line(0,1){1}}
\multiput(57.625,32.25)(0,2){11}{\line(0,1){1}}
\multiput(4.75,25.5)(2,0){4}{\line(1,0){1}}
\multiput(4.25,54.25)(2,0){2}{\line(1,0){1}}
\multiput(51.125,25.25)(2,0){6}{\line(1,0){1}}
\multiput(51.25,54.875)(2,0){3}{\line(1,0){1}}
\end{picture}
\caption{The mean entanglement entropy as a function of $n$ for $N = 60$, $z = 100$ and various $N_{\textrm{sq}}$. The continuous line is the approximation given by equation \eqref{eq:plateau}.}
\label{fig:plateau_massless_1d_ncr}
\end{figure}

In the previous sections \ref{subsec:results_massless} and \ref{subsec:results_mass}, where all modes were squeezed with the same parameter $z$, we showed that for large values of $z$ the mean entanglement entropy is well approximated by the large-squeezing expansion \cite{Katsinis:2023hqn}. In this regime, the mean entanglement entropy obeys a volume law, $\bar{S}_{\mathrm{EE}} \sim n$. For the state considered here, this conclusion is partly modified. As shown in figure \ref{fig:plateau_massless_1d}, the mean entanglement entropy has a simple form for large values of $z$. This form emerges for values of $z$ of the same order of magnitude as those that are required to reach the large-squeezing approximation when all modes are squeezed. It consists of a linear part for small subsystems, i.e. small values of $n$, followed by a constant part, i.e. an entanglement plateau, for large subsystems. The linear part corresponds to exactly the same volume law that is given by the large-squeezing approximation, i.e. $\bar{S}_{\mathrm{EE}} = z n$. The critical size of the subsystem $n_{\textrm{cr}}$, where the transition from the volume law to the entanglement plateau occurs, turns out to be equal to half the number of squeezed modes, as shown in figure \ref{fig:plateau_massless_1d_ncr}:
\begin{equation}
n_{\textrm{cr}} = \frac{N_{\textrm{sq}}}{2} .
\end{equation}

The mean entanglement entropy for large values of $z$ is well-approximated by the expression
\begin{equation}
\bar{S}_{\textrm{EE}} \simeq \begin{cases}
z n , & n \leq \frac{N_{\textrm{sq}}}{2} , \\
z \frac{N_{\textrm{sq}}}{2} , & \frac{N_{\textrm{sq}}}{2} \leq n \leq \frac{N}{2} .
\end{cases}
\label{eq:plateau}
\end{equation}
We present an analytic proof of this equation is section \ref{sec:plateau_theory}. We note that this expression includes only the dominant effect in the large-squeezing regime. Subleading terms of order $z^0$ that could result in the known logarithmic dependence are neglected. The entanglement entropy of this sector is obviously symmetric under the operation $n \leftrightarrow N - n$, implying that for arbitrary $n$, the complete expression is
\begin{equation}
\bar{S}_{\textrm{EE}} \simeq z \min \left( n , N - n , \frac{N_{\textrm{sq}}}{2} \right) .
\label{eq:plateau_min_expression}
\end{equation}
In order to recover a volume law for all $n$, all modes have to be squeezed. Then, the above formula degenerates to equation \eqref{eq:large_sq_one_d}.

\begin{figure}[ht]
\centering
\begin{picture}(92,31.5)
\put(0,0){\includegraphics[angle=0,width=0.9\textwidth]{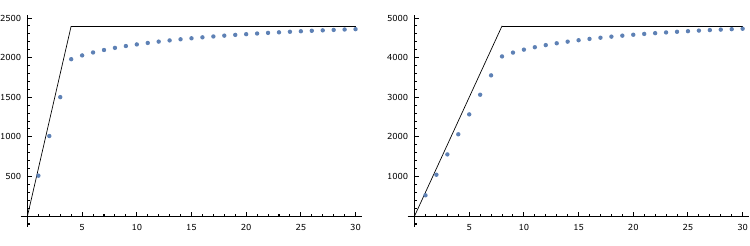}}
\put(44,3.25){{\footnotesize $n$}}
\put(90.375,3){{\footnotesize $n$}}
\put(2.5,28.125){{\footnotesize $\bar{S}_{\mathrm{EE} , 20}$}}
\put(49,27.875){{\footnotesize $\bar{S}_{\mathrm{EE} , 20}$}}
\put(19.5,28){{\small $N_{\textrm{sq}}=8$}}
\put(67.5,28){{\small $N_{\textrm{sq}}=16$}}
\multiput(8.625,4.5)(0,2){11}{\line(0,1){1}}
\multiput(60.375,4.5)(0,2){11}{\line(0,1){1}}
\multiput(4.5,26.375)(2,0){2}{\line(1,0){1}}
\multiput(50.875,26.375)(2,0){5}{\line(1,0){1}}
\end{picture}
\caption{The contribution to the mean entanglement entropy of the $\ell = 20$ sector, as a function of $n$ for $N = 60$, $z = 600$ and various $N_{\textrm{sq}}$. The continuous line is the approximation given by equation \eqref{eq:plateau}.}
\label{fig:plateau_massless_l20}
\end{figure}
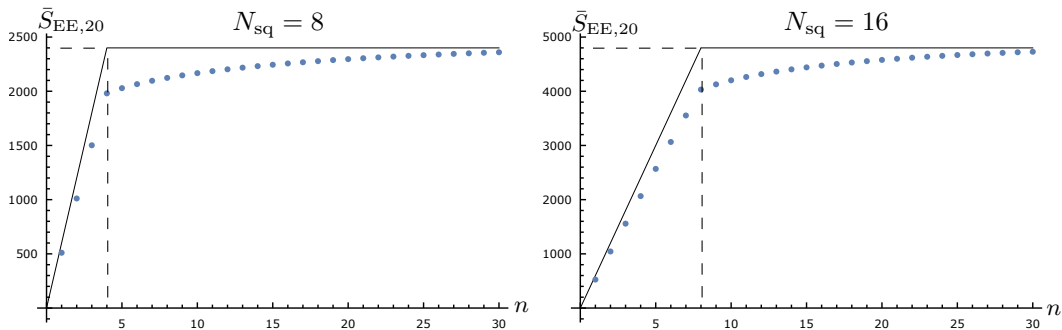

It turns out that the behaviour of all $\ell$-sectors with $\ell>0$ is similar to the behaviour of the $\ell=0$ sector. For large values of $z$, all $\ell$-sector contributions to the mean entanglement entropy are well-approximated by equation \eqref{eq:plateau}, as shown in figure \ref{fig:plateau_massless_l20} for $\ell=20$.
The only difference is that equation \eqref{eq:plateau} is a good approximation of the contribution of an $\ell$-sector to the mean entanglement entropy for values of $z$ larger than a critical value which is an increasing function of $\ell$, similarly to what happens when all modes are squeezed, as shown in figure \ref{fig:massless_sectors_smallz}. Therefore, we expect that the full mean entanglement entropy in the three-dimensional field theory for large $z$ is the sum of contributions of the form \eqref{eq:plateau}, one for each $\ell$-sector, with a different $N_{\textrm{sq}}(\ell)$. Using the $\ell_{\max}\sim n$ regularization, which is described in section \ref{sec:regularization}, the mean entanglement entropy is given by
\begin{equation}
\bar{S}_{\textrm{EE}} \simeq z \sum\limits_{\ell = 0}^{\left\lfloor c n \right\rfloor} \left( 2 \ell + 1 \right) \min \left( n , \frac{N_{\textrm{sq}}(\ell)}{2} \right) ,
\label{eq:plateau3d}
\end{equation}
where $c$ is the parameter of equation \eqref{eq:lmax}.

Let us consider a scenario in which $N_{\textrm{sq}}$ modes of each $\ell-$sector are squeezed with the same squeezing parameter that is sufficiently large for equation \eqref{eq:plateau} to apply. It is then obvious that the entanglement entropy obeys the standard large-squeezing volume law for $n< N_{\textrm{sq}}/2$, while it reaches a plateau for larger $n$. 

\begin{figure}[ht]
\centering
\begin{picture}(92,32)
\put(0,0){\includegraphics[angle=0,width=0.9\textwidth]{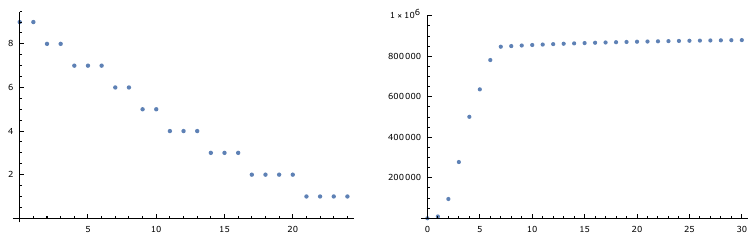}}
\put(43,2.75){{\footnotesize $\ell$}}
\put(90.375,2.75){{\footnotesize $n$}}
\put(1.5,28.75){{\footnotesize $N_{\textrm{sq}}$}}
\put(50.625,28.125){{\footnotesize $\bar{S}_{\mathrm{EE}}$}}
\end{picture}
\caption{The mean entanglement entropy as a function of $n$ for $N = 60$. All modes with frequencies smaller than one-fourth the maximal eigenfrequency of the $\ell = 0$ sector have been squeezed with $z = 1000$, while the rest lie at their ground state.}
\label{fig:plateau_3d_num}
\end{figure}
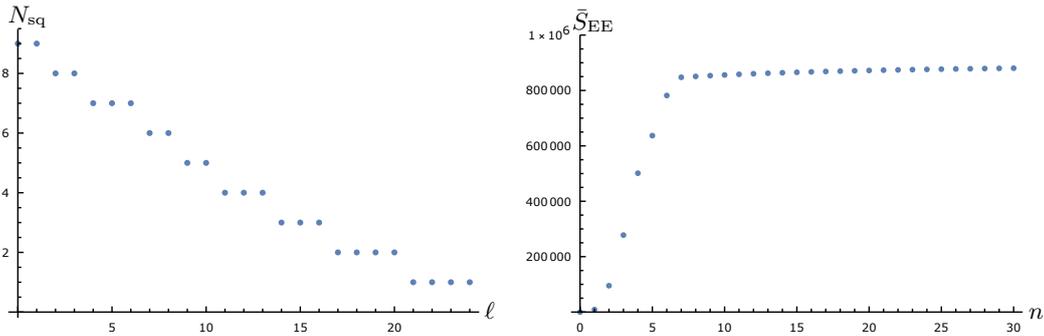

In an alternative scenario one may assume that all modes with eigenfrequencies below a critical value $\omega_{\textrm{cr}}$ are squeezed. This constraint implicitly defines the function $N_{\textrm{sq}}(\ell)$. For example, taking $\omega_{\textrm{cr}}$  equal to one-fourth the maximal eigenfrequency of the $\ell=0$ sector, the function $N_{\textrm{sq}}(\ell)$ and the corresponding entanglement entropy are shown in figure \ref{fig:plateau_3d_num}. We have checked both numerically and analytically that a volume term appears in the entanglement entropy before the plateau.

The appearance of a volume term is expected even for non-uniform squeezing of a subset of modes. We do not embark on an extensive analysis of the general case here. Instead, we focus on the $\ell=0$ sector, allowing for an arbitrary squeezing of its normal modes. 

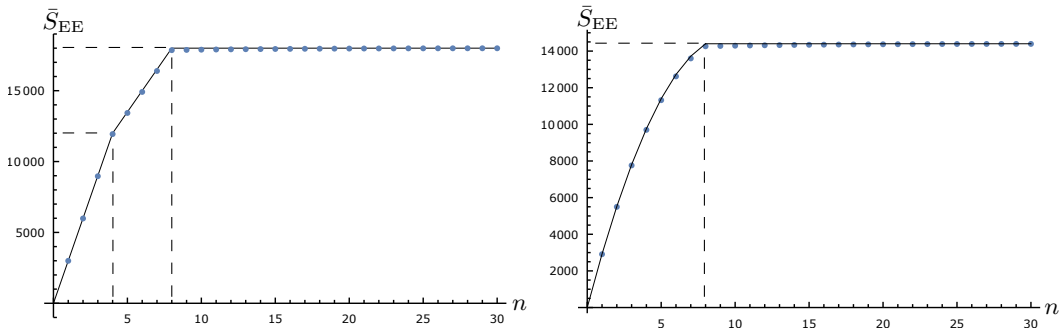
\begin{figure}[ht]
\centering
\begin{picture}(92,31.5)
\put(0,0){\includegraphics[angle=0,width=0.9\textwidth]{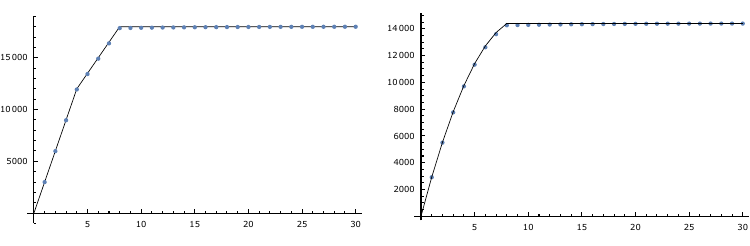}}
\put(44,3.375){{\footnotesize $n$}}
\put(90.375,3){{\footnotesize $n$}}
\put(3.125,28){{\footnotesize $\bar{S}_{\mathrm{EE}}$}}
\put(49.625,28.25){{\footnotesize $\bar{S}_{\mathrm{EE}}$}}
\multiput(9.25,4.5)(0,2){7}{\line(0,1){1}}
\multiput(14.375,4.5)(0,2){11}{\line(0,1){1}}
\multiput(60.75,4.5)(0,2){11}{\line(0,1){1}}
\multiput(4.875,26.375)(2,0){5}{\line(1,0){1}}
\multiput(4.875,18.875)(2,0){2}{\line(1,0){1}}
\multiput(51.25,26.75)(2,0){5}{\line(1,0){1}}
\end{picture}
\caption{The mean entanglement entropy as a function of $n$ for $N = 60$ and $N_{\textrm{sq}} = 16$. On the left panel $z_i = 3000$ for $1 \leq i \leq 8$ and $z_i = 1500$ for $9 \leq i \leq 16$. On the right panel $z_i = 3160 - 160 i$ for $1\leq i\leq 16$. The continuous line is the approximation given by equation \eqref{eq:plateau_arbitrary}.}
\label{fig:plateau_massless_random}
\end{figure}

Let $z_i$ be the squeezing parameter of the $i$-th mode, $Z$ the set $Z = \left\{ z_1 , z_2 , \ldots , z_{N_{\textrm{sq}}} \right\}$ and $\tilde Z$ the ordered permutation of $Z$, i.e. $\tilde{Z} = \left\{ \zeta_1 , \zeta_2 , \ldots , \zeta_{N_{\textrm{sq}}} \right\}$, so that $\zeta_1 > \zeta_2 > \ldots > \zeta_{N_{\textrm{sq}}}$. It can be shown numerically and analytically that the entanglement entropy is given by
\begin{equation}
\bar{S}_{\textrm{EE}} \simeq \begin{cases}
\frac{1}{2} \sum\limits_{i = 1}^{2 n} \zeta_i , & n \leq \frac{N_{\textrm{sq}}}{2} , \\
\frac{1}{2} \sum\limits_{i = 1}^{N_{\textrm{sq}}} \zeta_i , & \frac{N_{\textrm{sq}}}{2} \leq n \leq \frac{N}{2} .
\end{cases}
\label{eq:plateau_arbitrary}
\end{equation}
figure \ref{fig:plateau_massless_random} shows that this curve is indeed a good approximation of the mean entanglement entropy.

Non-uniform squeezing occurs naturally for the modes of scalar fields in curved gravitational backgrounds, such as cosmological spacetimes. The emergence of a volume term is a typical feature in these settings, even though noticeable exceptions may exist, such as for an exact de Sitter background \cite{Boutivas:2023ksg,Boutivas:2023mfg}.  

\section{Analytic derivation of the entanglement entropy for large squeezing of a subset of modes}
\label{sec:plateau_theory}

In section \ref{subsec:results_plateau} we studied states in which only some of the normal modes have been squeezed with the same squeezing parameter $z$, while the rest are in their ground state. We showed numerically that, in the limit that $z$ gets very large, the entanglement entropy assumes a very simple form, namely that of equation \eqref{eq:plateau}. In this section we derive this formula analytically. In order to do so, we employ the covariance matrix method \cite{Sorkin:2012sn}. It can be shown that this method is equivalent to the one we reviewed in section \ref{sec:review} and used in the rest of the paper, see appendix E of \cite{Katsinis:2023hqn}.

In the covariance matrix method we consider the $\left( 2 n \right) \times \left( 2 n \right)$ matrix $\mathcal{M}$ given by
\begin{equation}
\mathcal{M} = 2 i J \,\mathrm{Re}  M,
\end{equation}
where the covariance matrix  $M$ and the matrix $J$ are defined as
\begin{equation}
M = \begin{pmatrix}
\left\langle x_i x_j \right\rangle & \left\langle x_i \pi_j\right\rangle \\
\left\langle x_i \pi_j \right\rangle^T & \left\langle \pi_i \pi_j \right\rangle
\end{pmatrix} , \quad J = \begin{pmatrix}
0 & I \\
-I & 0
\end{pmatrix} .
\end{equation}
For the general Gaussian state \eqref{eq:squeezed_N} the matrix $\mathcal{M}$ reads
\begin{equation}
\mathcal{M}=i\begin{pmatrix}
- \left( \im W \left(\re W \right)^{-1} \right)_A & \left( \re W + \im W \left( \re W \right)^{-1} \im W \right)_A \\
- \left( \left( \re W \right)^{-1} \right)_A & \left( \left( \re W \right)^{-1} \im W \right)_A
\end{pmatrix} .
\end{equation}
The eigenvalues of $\mathcal{M}$ come in pairs of the form $(\lambda_i,-\lambda_i)$. The entanglement entropy is given by
\begin{equation}
S=\sum_{i=1}^{n} \left( \frac{\lambda_i + 1}{2} \ln \frac{\lambda_i + 1}{2} - \frac{\lambda_i - 1}{2} \ln\frac{\lambda_i - 1}{2} \right) ,
\label{eq:S_of_M}
\end{equation}
where $\lambda_i$ are the positive eigenvalues of $\mathcal{M}$.

If we consider a general squeezed state, where the $i$-th mode has been squeezed with a squeezing parameter $z_i$, the matrix $\mathcal{M}$ reads
\begin{equation}
\mathcal{M} = i \begin{pmatrix}
\mathcal{M}_{XP} & \mathcal{M}_{PP} \\
- \mathcal{M}_{XX} & - \mathcal{M}_{XP}
\end{pmatrix} ,
\end{equation}
where
\begin{align}
\mathcal{M}_{XX} &= \sum_{i = 1}^N \frac{\cosh z_i + \sinh z_i \sin \left( 2 \omega_i t + \phi_i \right)}{\omega_i} \left( v_i \right)_A \left( v_i \right)^T_A , \\
\mathcal{M}_{PP} &= \sum_{i = 1}^N \omega_i \left( \cosh z_i - \sinh z_i \sin \left( 2 \omega_i t + \phi_i \right) \right) \left( v_i \right)_A \left( v_i \right)^T_A , \\
\mathcal{M}_{XP} &= \sum_{i = 1}^N \sinh z_i \cos \left( 2 \omega_i t + \phi_i \right) \left( v_i \right)_A \left( v_i \right)^T_A .
\end{align}
The vector $v_i$ is the normalized shape vector of the $i$-th mode, i.e. when only the $i$-th mode has been excited, the classical motion of the system is $x_j = A \left( v_i \right)_j \cos \left( \omega_i t + \phi_i \right)$. The vector $\left( v_i \right)_A$ is the block of $v_i$ that contains its first $n$ components.

We consider a state in which only $N_{\mathrm{sq}}$ modes are squeezed with the same squeezing parameter $z$. Without loss of generality, we consider that these are the modes with indices $i = 1 , 2 , \ldots , N_{\textrm{sq}}$. Then, one can express the matrix $\mathcal{M}$ as
\begin{equation}
\mathcal{M}= e^z \mathcal{M}^{(+)} + e^{-z} \mathcal{M}^{(-)} + \mathcal{M}^{(0)} .
\end{equation}
In an obvious manner, in the limit that $z$ becomes very large, the behaviour of the system is determined by the matrix $\mathcal{M}^{(+)}$. In this limit we expect that some of the eigenvalues of $\mathcal{M}$ will be of order $e^z$. More specifically these eigenvalues will be equal to $\lambda_i \simeq \lambda_i^{(+)} e^z$, where $\lambda_i^{(+)}$ are the non-vanishing eigenvalues of the matrix $\mathcal{M}^{(+)}$. When an eigenvalue $\lambda$ of the matrix $\mathcal{M}$ is very large, its contribution to the entanglement entropy can be approximated as,
\begin{equation}
\frac{\lambda + 1}{2} \ln \frac{\lambda + 1}{2} - \frac{\lambda - 1}{2} \ln\frac{\lambda - 1}{2} = \ln\frac{\lambda}{2} + \ldots
\end{equation}
As a result, for large $z$ we obtain
\begin{equation}
S = z N^{(+)} + \ldots ,
\label{eq:S_plateau_largez}
\end{equation}
where $N^{(+)}$ is the number of positive eigenvalues of the matrix $\mathcal{M}^{(+)}$.

The matrix $\mathcal{M}^{(+)}$ reads
\begin{equation}
\mathcal{M}^{(+)} = i \begin{pmatrix}
\mathcal{M}_{XP}^{(+)} & \mathcal{M}_{PP}^{(+)} \\
- \mathcal{M}_{XX}^{(+)} & - \mathcal{M}_{XP}^{(+)}
\end{pmatrix} ,
\end{equation}
where
\begin{align}
\mathcal{M}_{XX}^{(+)} &= \frac{1}{2} \sum_{i = 1}^{N_{\textrm{sq}}} \frac{1 + \sin \left( 2 \omega_i t + \phi_i \right)}{\omega_i} \left( v_i \right)_A \left( v_i \right)^T_A , \\
\mathcal{M}_{PP}^{(+)} &= \frac{1}{2} \sum_{i = 1}^{N_{\textrm{sq}}} \omega_i \left( 1 - \sin \left( 2 \omega_i t + \phi_i \right) \right) \left( v_i \right)_A \left( v_i \right)^T_A , \\
\mathcal{M}_{XP}^{(+)} &= \frac{1}{2} \sum_{i = 1}^{N_{\textrm{sq}}} \cos \left( 2 \omega_i t + \phi_i \right) \left( v_i \right)_A \left( v_i \right)^T_A .
\end{align}
The matrix can be written as
\begin{equation}
\mathcal{M}^{(+)} = \begin{pmatrix}
O^T & 0 \\
0 & O^T
\end{pmatrix} \tilde{\mathcal{M}}^{(+)} \begin{pmatrix}
O & 0 \\
0 & O
\end{pmatrix} .
\label{eq:Mplustilde}
\end{equation}
$\tilde{\mathcal{M}}^{(+)}$ is an $\left( 2 N_{\textrm{sq}} \right) \times \left( 2 N_{\textrm{sq}} \right)$ square matrix that reads
\begin{equation}
\tilde{\mathcal{M}}^{(+)} = i \begin{pmatrix}
\tilde{\mathcal{M}}_{XP}^{(+)} & \tilde{\mathcal{M}}_{PP}^{(+)} \\
- \tilde{\mathcal{M}}_{XX}^{(+)} & - \tilde{\mathcal{M}}_{XP}^{(+)}
\end{pmatrix} ,
\end{equation}
where $\tilde{\mathcal{M}}_{XX}^{(+)}$, $\tilde{\mathcal{M}}_{PP}^{(+)}$ and $\tilde{\mathcal{M}}_{XP}^{(+)}$ are diagonal matrices with elements
\begin{align}
\left( \tilde{\mathcal{M}}_{XX}^{(+)} \right)_{ij} &= \frac{1}{2} \frac{1 + \sin \left( 2 \omega_i t + \phi_i \right)}{\omega_i} \delta_{ij} , \\
\left(\tilde{\mathcal{M}}_{PP}^{(+)} \right)_{ij}&= \frac{1}{2} \omega_i \left( 1 - \sin \left( 2 \omega_i t + \phi_i \right) \right) \delta_{ij} , \\
\left(\tilde{\mathcal{M}}_{XP}^{(+)} \right)_{ij}&= \frac{1}{2} \cos \left( 2 \omega_i t + \phi_i \right) \delta_{ij} .
\end{align}
$O$ is the $N_{\textrm{sq}}\times n$ matrix whose elements read
\begin{equation}
O_{ij} = \left( v_i \right)_j .
\end{equation}
It can be shown that the matrix $\tilde{\mathcal{M}}^{(+)}$ has $2 N_{\textrm{sq}}$ vanishing eigenvalues, but only $N_{\textrm{sq}}$ eigenvectors.

Let us consider the eigenvalue problem for the matrix $\mathcal{M}^{(+)}$. This requires the solution of the algebraic equation
\begin{equation}
\det \left( \lambda I_{2 n} - \mathcal{M}^{(+)} \right) = 0 .
\end{equation}
Since we are interested in the non-vanishing eigenvalues of $\mathcal{M}^{(+)}$, we may divide this equation with $\lambda$ to get
\begin{equation}
\det \left( I_{2 n} - \frac{1}{\lambda} \mathcal{M}^{(+)} \right) = 0 .
\end{equation}
Sylvester's determinant formula states that $\det \left( I_r + A B \right) = \det \left( I_s + B A \right)$, where $A$ is an $r \times s$ matrix and $B$ is an $s \times r$ matrix. Employing this formula and equation \eqref{eq:Mplustilde} allows the re-expression of the above as
\begin{equation}
\det \left( I_{2 N_{\textrm{sq}}} - \frac{1}{\lambda} \hat{\mathcal{M}}^{(+)} \right) = 0 ,
\end{equation}
where $\hat{\mathcal{M}}^{(+)}$ is a $\left( 2 N_{\textrm{sq}} \right) \times \left( 2 N_{\textrm{sq}} \right)$ square matrix that reads
\begin{equation}\label{eq:factorization}
\hat{\mathcal{M}}^{(+)} = \begin{pmatrix}
OO^T & 0 \\
0 & OO^T
\end{pmatrix} \tilde{\mathcal{M}}_+ .
\end{equation}
Sylvester's determinant formula implies that $\hat{\mathcal{M}}^{(+)}$ and $\mathcal{M}^{(+)}$ share the same non-vanishing eigenvalues. Moreover due to the form of equation \eqref{eq:factorization}, the matrix $\hat{\mathcal{M}}^{(+)}$ has at least $N_{\textrm{sq}}$ vanishing eigenvalues that correspond to the eigenvectors of the matrix $\tilde{\mathcal{M}}^{(+)}$.

Recall that the eigenvalues of matrix $\mathcal{M}$ come in pairs of opposite eigenvalues. Naturally this must hold for the eigenvalues that are of order $e^z$, and, thus, for the eigenvalues of $\mathcal{M}^{(+)}$. The maximal number of non-vanishing eigenvalues of $\mathcal{M}^{(+)}$ is $2 n$. On the other hand the matrix $\hat{\mathcal{M}}^{(+)}$ has at least $N_{\textrm{sq}}$ vanishing eigenvalues, and, thus, at most $N_{\textrm{sq}}$ non-vanishing ones. As $\hat{\mathcal{M}}^{(+)}$ and $\mathcal{M}^{(+)}$ have the same non-vanishing eigenvalues, it follows that $\mathcal{M}^{(+)}$ has at most $\min\left( 2 n , N_{\textrm{sq}} \right)$ non-vanishing eigenvalues. Since half of them will be positive, this implies that the entanglement entropy for large $z$, as follows from equation \eqref{eq:S_plateau_largez}, equals\footnote{When $N_{\textrm{sq}}$ is odd $\mathcal{M}$ has $\lfloor N_{\textrm{sq}}/2 \rfloor$ eigenvalues of order $e^z$ and one of order $e^{z / 2}$, making the equation \eqref{eq:plateau_theory} correct for odd $N_{\textrm{sq}}$ as well.}
\begin{equation}
S = z \min \left( n , N_{\textrm{sq}}/2 \right) + \ldots ,
\label{eq:plateau_theory}
\end{equation}
which is identical to the formula \eqref{eq:plateau} that we obtained numerically.

Strictly speaking, we showed that the above formula is an upper bound for the large-$z$ behaviour of entanglement entropy in an arbitrary harmonic system where a random subset of $N_{\textrm{sq}}$ harmonic modes have been squeezed with the same parameter $z$,\footnote{The same idea may be applied when the squeezing parameters are different for each mode. One should start by specifying how many non-vanishing eigenvalues of the covariance matrix exist at order $e^{z_{\max}}$, where $z_{\max}$ is the largest of the squeezing parameters, and then, advance to the second largest one and so on. Such treatment should provide the formula \eqref{eq:plateau_arbitrary} that we obtained numerically.} whereas the rest remain at their ground state. Our numerical results in section \ref{subsec:results_plateau} indicate that this bound is actually saturated for large $z$. Although in section \ref{subsec:results_plateau} the squeezed modes were selected to be the lowest eigenfrequency modes, the large $z$ behaviour is independent of this choice.

The details of the coupling matrix of the harmonic system have not been taken into account in this section. The specifics of the coupling matrix may lead to more vanishing eigenvalues of the covariance matrix at order $e^z$. This is the case when the harmonic system can be divided into non-interacting sectors, like the three-dimensional field theory that was studied numerically in section \ref{subsec:results_plateau}. In such cases, the upper bound for the large-$z$ behaviour of entanglement entropy has to be applied to the contribution of each sector separately.

\section{Conclusions}
\label{sec:discussion}

We studied entanglement entropy in free scalar field theory lying in a squeezed state in $3 + 1$ dimensions. We followed the approach of \cite{srednicki} that employs a discretized version of the theory, summarized in section \ref{subsec:discretization}. For spherical smooth entangling surfaces, which are our main point of interest, it is difficult to introduce a discretization scheme that respects the symmetries of the entangling surface and gives rise to a homogeneous density of the degrees of freedom. We resolved this problem through the introduction of an appropriate regularization scheme for the calculation of entanglement entropy, discussed in section \ref{subsec:reg_lmax}. Instead of studying the entanglement entropy at a given time, we study the mean entanglement entropy over time, which preserves the symmetries of the Hamiltonian for the states that we consider, as we show in the appendix.

We found that squeezing introduces a volume term in the entanglement entropy. This is expected for a squeezed state \cite{Katsinis:2023hqn}, but more importantly it is expected for an arbitrary quantum state \cite{Page:1993df}. This finding implies that the area law for the entanglement entropy in field theory, found in \cite{srednicki}, is not a general property, but rather a very special one that appears in the ground state or in a coherent state.

Our calculation indicates that states of a harmonic system in which all normal modes lie 
in a squeezed configuration display the behaviour expected from an arbitrary quantum state in the
context of Page's argument \cite{Page:1993df}. According to the latter, the bipartite entanglement entropy in an arbitrary state is close to maximal and proportional to the volume of the smaller subsystem. Therefore, the squeezed states that we studied can serve as a benchmark tool for the investigation of general properties of entanglement.

Page's calculation implies that the proportionality constant that connects the volume of the smaller subsystem to the entanglement entropy depends on the dimensionality of the Hilbert space of the local degrees of freedom. In our case this space is infinite-dimensional. Obviously the constant that we calculate is not infinite. For large squeezing the constant is equal to the squeezing parameter, as implied by the large-squeezing expansion \cite{Katsinis:2023hqn} that results in equation (\ref{eq:large_sq_lmax_reg}) for the entropy. This means that the squeezing parameter effectively acts as the dimension of the Hilbert space. It measures how many states in the Hilbert space of the local degrees of freedom can contribute significantly
to long-range entanglement resulting in a volume effect.

Another interesting outcome of our calculation concerns the effect of the field
mass on the entanglement entropy. It is well known that mass suppresses entanglement in the ground state of the field theory. However, the enhancement of entanglement entropy by the squeezing of the state appears to be independent of the field mass in sufficiently squeezed states.

When only a subset of the normal modes lie in a squeezed state, the behaviour of entanglement entropy is more complicated. In $(1 + 1)$-dimensional field theory and for large squeezing parameters, the volume law emerges again, but for subsystems smaller than a critical size. For subsystems larger than this critical size, the entanglement entropy becomes independent of the size of the subsystem, reaching a plateau. A subsystem with the critical size has a number of degrees of freedom equal to half the number of the squeezed modes.

This behaviour is consistent with the following pictorial interpretation: Squeezed modes act as threads of entanglement that connect the two subsystems. Each of these modes gives rise to entanglement entropy equal to half the corresponding squeezing parameter. We can imagine that entanglement entropy stems from these threads, but under the constraint that each degree of freedom of one subsystem can be connected to only two threads, a picture reminiscent of the idea of entanglement monogamy. By giving priority to the threads of maximal squeezing, the assignment of the threads to the degrees of freedom is such that the generated entanglement entropy is the maximal possible.

In $(3 + 1)$-dimensional field theory, the entanglement entropy is the sum of the contributions of the decoupled angular momentum sectors. Each of those has a similar behaviour to a $(1 + 1)$-dimensional field theory. Therefore the behaviour of the overall entanglement entropy depends on the number of modes that have been squeezed in each sector. It turns out that the overall behaviour is similar to the $(1 + 1)$-dimensional case. Entanglement entropy is an increasing function of the size of the entangling sphere up to a critical size. For larger entangling spheres it reaches an approximate plateau. In the regime that the entanglement entropy is an increasing function of the radius of the entangling sphere, a volume term is present, independently of the number of squeezed modes. This is consistent with the fact that, for random Gaussian states, the spectra of the reduced density matrices concentrate around that of a thermal state \cite{randomGaussian}.

The area law property of the entanglement entropy of field theory in its ground state resembles the famous property of black hole entropy. This resemblance has inspired scenarios of an interpretation of gravity as an entropic force. Our findings imply that, in such scenarios, the underlying degrees of freedom that give rise to entropic gravity should lie in their ground state, or at least close to it. 

\acknowledgments

The research of N. Tetradis was supported by the Hellenic Foundation for Research and Innovation (H.F.R.I.) under the “First Call for H.F.R.I. Research Projects to support Faculty members and Researchers and the procurement of high-cost research equipment grant” (Project Number: 824). The research of D. Katsinis was supported by the FAPESP Grant No. 2021/01819-0.

\appendix

\section{The state and the symmetries of the system}
\label{app:state}

In this appendix we consider a harmonic system with all its modes lying in a squeezed state with the same squeezing parameter. We show that in such a state the squares of the uncertainties of the local degrees of freedom and their conjugate momenta vary in time. Nevertheless, their mean over time preserves the symmetries of the harmonic system.

Let us recall the squeezed state \eqref{eq:squeezed}. An oscillator at this state satisfies
\begin{align}
\left< x \right> = x_0 \left( t \right) , &\quad \left< x^2 \right> = x_0^2 \left( t \right) + \frac{\hbar}{2 m \omega} \left( \cosh z + \sinh z \sin 2 \omega \left( t - t_0 \right) \right) , \\
\left< p \right> = p_0 \left( t \right) , &\quad \left< p^2 \right> = p_0^2 \left( t \right) + \frac{\hbar m \omega}{2} \left( \cosh z - \sinh z \sin 2 \omega \left( t - t_0 \right) \right) ,
\end{align}
implying that
\begin{align}
\left( \delta x \right)^2 &= \frac{\hbar}{2 m \omega} \left( \cosh z + \sinh z \sin 2 \omega \left( t - t_0 \right) \right) , \\
\left( \delta p \right)^2 &= \frac{\hbar m \omega}{2} \left( \cosh z - \sinh z \sin 2 \omega \left( t - t_0 \right) \right) .
\end{align}
We denote the mean in time with a bar. Since ${x}_0 \left( t \right)$, ${p}_0 \left( t \right)$ are given by \eqref{eq:coherent_mean_x} and \eqref{eq:coherent_mean_p}, the above equations imply that
\begin{align}
\overline{\left< x \right>} = 0 , &\quad \overline{\left< x^2 \right>} = \frac{x_0^2}{2} + \frac{\hbar}{2 m \omega} \cosh z , \\
\overline{\left< p \right>} = 0 , &\quad \overline{\left< p^2 \right>} = \frac{p_0^2}{2} + \frac{\hbar m \omega}{2} \cosh z , \\
\overline{\left( \delta x \right)^2} = \frac{\hbar}{2 m \omega} \cosh z , &\quad
\overline{\left( \delta p \right)^2} = \frac{\hbar m \omega}{2} \cosh z .
\end{align}

Let $O$ be the orthogonal transformation that connects the normal coordinates $\tilde{x}_i$ to the local coordinates $x_i$,
\begin{equation}
x_i = \sum_j O_{ij} \tilde{x}_j , \quad p_i = \sum_j O_{ij} \tilde{p}_j .
\end{equation}
We assume that each mode lies in a squeezed state with squeezing parameter $z_i$. Then,
\begin{align}
\left< x_i \right> 
&= \sum_j O_{ij} \tilde{x}_{0 j} \left( t \right), \\
\left< x_i^2 \right>
&= \left( \sum_j O_{ij} \tilde{x}_{0 j} \left( t \right) \right)^2 + \frac{\hbar}{2 m} \sum_j O_{ij} O_{ij} \frac{1}{\omega_j} \left( \cosh z_j + \sinh z_j \sin 2 \omega \left( t - t_{0j} \right) \right) , \\
\left< p_i \right>
&= \sum_j O_{ij} \tilde{p}_{0 j} \left( t \right) , \\ 
\left< p_i^2 \right>
&= \left( \sum_j O_{ij} \tilde{p}_{0 j} \left( t \right) \right)^2 + \frac{\hbar m}{2} \sum_j O_{ij} O_{ij} \omega_j \left( \cosh z_j - \sinh z_j \sin 2 \omega \left( t - t_{0j} \right) \right) ,
\end{align}
which directly imply that
\begin{align}
\left( \delta x_i \right)^2 &= \frac{\hbar}{2 m} \sum_j O_{ij} O_{ij} \frac{1}{\omega_j} \left( \cosh z_j + \sinh z_j \sin 2 \omega \left( t - t_{0j} \right) \right) , \\
\left( \delta p_i \right)^2 &= \frac{\hbar m}{2} \sum_j O_{ij} O_{ij} \omega_j \left( \cosh z_j - \sinh z_j \sin 2 \omega \left( t - t_{0j} \right) \right) .
\end{align}
Therefore,
\begin{align}
\overline{\left< x_i \right>} = 0 , &\quad \overline{\left< x_i^2 \right>} = \sum_j O_{ij} \frac{\tilde{x}_{0 j}^2}{2} + \frac{\hbar}{2 m} \sum_j O_{ij} O_{ij} \frac{1}{\omega_j} \cosh z_j , \\
\overline{\left< p_i \right>} = 0 , &\quad \overline{\left< p_i^2 \right>} = \sum_j O_{ij} \frac{\tilde{p}_{0 j}^2}{2} + \frac{\hbar m}{2} \sum_j O_{ij} O_{ij} \omega_j \cosh z_j
\end{align}
and
\begin{equation}
\overline{\left( \delta x_i \right)^2} = \frac{\hbar}{2 m} \sum_j O_{ij} O_{ij} \frac{1}{\omega_j} \cosh z_j , \quad \overline{\left( \delta p_i \right)^2} = \frac{\hbar m}{2} \sum_j O_{ij} O_{ij} \omega_j \cosh z_j .
\end{equation}

When all modes are squeezed with the same squeezing parameter $z$, the above yield
\begin{equation}
\overline{\left( \delta x_i \right)^2} = \frac{\hbar \cosh z}{2 m} \sum_j O_{ij} O_{ij} \frac{1}{\omega_j} , \quad \overline{\left( \delta p_i \right)^2} = \frac{\hbar m \cosh z}{2} \sum_j O_{ij} O_{ij} \omega_j .
\end{equation}
Bearing in mind that the eigenfrequency matrix $\tilde{\Omega}$ in the basis of the normal coordinates is $\tilde{\Omega}_{ij} = \omega_i \delta_{ij}$, the quantities $\sum_j O_{ij} O_{ij} \frac{1}{\omega_j}$ and $\sum_j O_{ij} O_{ij} \omega_j$ are simply the $ii$ components of $\Omega^{-1}$ and $\Omega$ as expressed in the local coordinates. Therefore
\begin{equation}
\overline{\left( \delta x_i \right)^2} = \frac{\hbar \cosh z}{2 m} \Omega^{-1}_{ii} , \quad \overline{\left( \delta p_i \right)^2} = \frac{\hbar m \cosh z}{2} \Omega_{ii} .
\end{equation}
The matrix $\Omega$ defines the dynamics of the harmonic system. For example, if we consider a free field theory, the matrix $\Omega$ is translationally and rotationally invariant. Therefore, it follows that the mean uncertainties in time, when we squeeze all modes with the same squeezing parameter, preserve the symmetries of the overall Hamiltonian.
Notice that in general the above argument still holds if the squeezing parameters are chosen so that they are a given function of the eigenfrequency of the corresponding mode.


\end{document}